\documentclass[twoside,11pt]{article}

\usepackage{blindtext}

\usepackage{jmlr2e}

\usepackage{amsmath}

\usepackage{amssymb}
\usepackage{xcolor,comment}
\usepackage{natbib} 
\usepackage{pdfpages}

\graphicspath{{./}}

\usepackage{enumerate}

\usepackage{hyperref, bm, lmodern}       
\usepackage{url}            

\newtheorem{prop}{Proposition}

\newtheorem{corollry}{Corollary}

\DeclareMathOperator{\tr}{tr}
\DeclareMathOperator{\Bias}{Bias}

\newcommand\numberthis{\addtocounter{equation}{1}\tag{\theequation}}

\allowdisplaybreaks 

\usepackage{lastpage}

\ShortHeadings{Inference for SGD}{Rahul Singh, Abhinek Shukla and Dootika Vats}
\firstpageno{1}

\begin{document}


\title{On the Utility of Equal Batch Sizes for Inference in  Stochastic Gradient Descent}

\author{\name Rahul Singh \email wrahulsingh@gmail.com \\
       \addr Department of Mathematics\\
       Indian Institute of Technology Delhi, India
       \AND
       \name Abhinek Shukla \email abhushukla@gmail.com \\
       \addr Department of Statistics and Data Science\\
       National University of Singapore, Singapore
       \AND
        \name Dootika Vats \email dootika@iitk.ac.in \\
       \addr Department of Mathematics and Statistics\\
       Indian Institute of Technology Kanpur, India
       }

\editor{ }

\maketitle

\thispagestyle{plain}


\begin{abstract}
Stochastic gradient descent (SGD) is an estimation tool for large data employed in machine learning and statistics. Due to the Markovian nature of the SGD process, inference is a challenging problem.  An underlying asymptotic normality of the averaged SGD (ASGD) estimator allows for the construction of a batch-means estimator of the asymptotic covariance matrix. Instead of the usual increasing batch-size strategy, we propose a memory efficient equal batch-size strategy and show that under mild conditions, the { batch-means} estimator is consistent. A key feature of the proposed batching technique is that it allows for bias-correction of the variance, at no { additional} cost to memory. {  Further}, since joint inference for large dimensional problems may be undesirable, we present marginal-friendly simultaneous confidence intervals, and show through an example { on} how covariance estimators of ASGD can be employed { for} improved predictions. 
\end{abstract}

	
\section{Introduction}

{ Stochastic gradient descent (SGD) is a popular and efficient optimization technique seminally introduced by \cite{rm1951}.} Given the nature of modern data, the increasing popularity of SGD is natural, owing to computational efficiency for large data-sets, and compatibility in online settings \citep[see, e.g., ][]{bottou2010,bottou,sgd2017}.
    
We assume data arise from $\Pi$, a probability distribution on $\mathbb{R}^{r}$, denoted by $\zeta \sim \Pi$. In a model fitting paradigm, a function $f:\mathbb{R}^{d} \times \mathbb{R}^{r} \to \mathbb{R}$ typically measures  empirical loss for estimating a parameter $\theta$, having observed the data, $\zeta$. Denote the expected loss as $F(\theta)=\mathbb{E}_{\zeta\sim\Pi}\left[f(\theta,\zeta)\right]$. The main parameter of interest is $\theta^*\in\mathbb{R}^d$ where
\begin{align}\label{problem}
\theta^*=\arg\min_{\theta\in\mathbb{R}^d}F(\theta).
\end{align}
With data $\zeta_i \overset{\text{iid}}{\sim} \Pi$ for $i = 1, \dots, n$, the goal is to estimate $\theta^*$, { where iid refers to independently and identically distributed}. This invariably involves a gradient based technique. Often, $F(\theta)$ is unavailable and a first approximation step is to replace $F(\theta)$ with the empirical loss, $n^{-1} \sum_{i=1}^{n} f(\theta, \zeta_i)$.  When the data are online or when calculation of the complete gradient vector is expensive, a further adjustment is made by replacing the complete gradient with an unbiased estimate. This yields a large class of stochastic gradient algorithms. Denote $\nabla f(\theta,\zeta)$ as the gradient vector of $f(\theta,\zeta)$ with respect to $\theta$, $\eta_i > 0$ as a  learning rate, and $\theta_0$ as the starting point of an SGD process. The $i^{\text{th}}$ iterate of  SGD is:
\begin{align}\label{iterate:sequence}
\theta_i=\theta_{i-1}-\eta_i\nabla f(\theta_{i-1},\zeta_i)\,, \qquad \text{ for } i = 1,2,\ldots\,\,\,.
\end{align} 
Despite the approximations introduced in the optimization, SGD estimates of $\theta$ can have nice statistical properties \citep{fabian,ruppert1988,pj1992}, particularly when $\eta_i$ is appropriately decreasing and the estimator of $\theta^*$ is chosen to be the averaged SGD (ASGD): 
\[
\hat \theta_n:=n^{-1}\sum_{i=1}^{n}\theta_i\,.
\] 
Naturally, a point estimate of $\theta^*$ alone is not sufficient. The work of \cite{pj1992} has particularly been instrumental in building a framework for statistical inference for $\hat{\theta}_n$. Let $A := \nabla^2 F(\theta^*)$ denote the Hessian of $F(\theta)$ evaluated at $\theta = \theta^*$ and define $S := \mathbb{E}_\Pi\left([\nabla f(\theta^*,\zeta)][\nabla f(\theta^*,\zeta)]^\top\right)$. When the derivative and expectation are interchangeable, $\mathbb{E}_{\Pi} \left[\nabla f(\theta^*,\zeta) \right] = \nabla F(\theta^*) = 0$. \cite{pj1992} showed that if $F$ is strictly convex with a Lipschitz gradient and $\eta_i=\eta i^{-\alpha}$ with $\alpha\in(0.5,1)$, then $\hat \theta_n$ is a consistent estimator of $\theta^*$, and under some additional conditions,
\begin{align}
\label{eq:pj_normal}
\sqrt{n}(\hat \theta_n-\theta^*)\xrightarrow{\texttt{d}} N(0,\Sigma)\text{ as }n\to\infty, \text{ where } \Sigma=A^{-1}SA^{-1}.
\end{align}

{   For a true end-to-end analysis, in addition to estimating $\theta^*$, a practitioner would be interested in assessing the quality of this estimator by estimating $\Sigma$, employing estimators of $\Sigma$ for inference, and equipping predictions with uncertainty estimates. Thus,} statistical inference for model parameters is a way forward towards robust implementations of machine learning algorithms. Although there is adequate literature devoted to the convergence behavior of the ASGD estimator and its variants \citep{zhang:2004,nami:2008,aggrawal:2012,dong2021},  estimators of $\Sigma$ have only recently been developed \citep{chen2020aos,chen2022online,sgd:bootstrap,leung:chan:2022,chen2021jasa}. Robustness and quality of inference depend critically on the quality of estimation of $\Sigma$.
	
\cite{chen2020aos} proposed two consistent estimators of $\Sigma$: an expensive plug-in estimator that requires repeated computation of the inverse of a Hessian, and a variant of the traditional batch-means estimator of \cite{chen1987multivariate} that is cheap to implement. In a batch-means estimator, SGD iterates are broken into batches of possibly differing sizes. A weighted sample covariance of the resulting batch-mean vectors yields a batch-means estimator; the quality of estimation is affected by the choice of batch-sizes. \cite{chen2021jasa} proposed a novel increasing batch-size strategy where the size of the batches continually increases until saturation, at which point a new batch is created.  We refer to this estimator as the increasing batch-size (IBS) estimator.

Despite the novel batching strategy, finite-sample performance for the IBS estimator is underwhelming; we demonstrate this in a variety of examples. A primary reason for the under-performance is that as iteration size increases, the  sample mean vectors of all but the last batch cannot improve in quality. We employ equal batch-sizes, that are carefully chosen for both practical utility and theoretical guarantees. Specifically, our proposed batch-sizes are powers of two, where the powers increase as a function of the iteration length. Under mild conditions, our batching strategy yields a consistent estimator and we obtain mean-square-error bounds. 

Equal batch-size (EBS) batch-means estimators are common-place in the Markov chain Monte Carlo (MCMC) literature \citep[see, e.g.][]{geyer1992,jones2006,flagel2010}. However, the Markov chains generated by MCMC and  SGD are fundamentally different; MCMC typically produces a time-homogeneous, stationary, ergodic chain and SGD produces a time-inhomogeneous and non-stationary chain that converges to a Dirac mass distribution \citep{bridging}. Due to these differences, the existing theoretical results of the batch-means estimator of MCMC are not applicable for SGD. However, as we will see, the tools utilized in output analysis for MCMC find use in setting up a workflow for statistical inference in SGD.

The proposed doubling batching structure is developed to allow for finite-sample improvements in the estimation of $\Sigma$. As discussed in \cite{chen2020aos,chen2021jasa}, due to the Markovian structure of $\{\theta_i\}_{i=1}^{n}$,  the estimator of $\Sigma$ is often under-biased for any finite $n$; this bias is one of the primary reasons for the underwhelming inferential performance of most estimators of $\Sigma$. Our batching technique allows for a memory-efficient, consistent, and a  bias-reduced estimator of $\Sigma$ using the lugsail technique of \cite{vats:lugsail}. Such a bias-reduction technique cannot be applied to the IBS estimator of \cite{chen2021jasa}.	 


{ 
It is worth considering how practitioners are expected to use estimators of $\Sigma$: two potential uses are as follows. First, estimating $\Sigma$ allows for an interpretation on the quality of estimation of $\theta^*$, particularly in smaller dimensions. Secondly, estimators of $\Sigma$ can then be employed to yield estimators of variability of functions of $\hat{\theta}_n$, aiding in uncertainty quantification for predictions. We explore this second point more carefully in Section~\ref{sec:sims}. 
}
    %
%
Further, using the asymptotic normality in \eqref{eq:pj_normal} and consistent estimators of $\Sigma$, it is possible to implement traditional multivariate hypothesis tests. { That is, a consistent estimator of $\Sigma$ can be used to construct an ellipsoidal confidence region for $\theta^*$ using \eqref{eq:pj_normal}. In high dimensional prediction models, where testing may not be a priority,  such confidence regions cease to convey a useful interpretation. Instead, simultaneous hyper-rectangular confidence regions allow easy interpretations for every marginal component, while also retaining coverage of the confidence region. This naturally, comes at the cost of the volume of the confidence region. We term such hyper-rectangular regions as ``marginal-friendly''.}  So far, estimators of $\Sigma$ have been employed to make either uncorrected marginal confidence intervals, or uninterpretable ellipsoidal confidence regions. Adapting tools developed in stochastic simulation, we construct { marginal-friendly confidence regions} with simultaneous coverage that utilize consistent estimators of $\Sigma$.

The rest of the paper is organized as follows. In Section~\ref{sec:BM_estimator} we present our proposed batching strategy. Assumptions and proof of consistency of the resulting batch-means estimator are in Section~\ref{sec:main:results}. Section~\ref{sec:lugsail} describes the under-estimation problem in estimating $\Sigma$, and discusses bias-correction through a lugsail estimator, for which  { we obtain the same rate of convergence as the original batch-means estimator}. Section~\ref{sec:regions} presents the structure of the marginal-friendly confidence regions. The performance of our proposed estimator is demonstrated through two simulated data problems in Section~\ref{sec:sims}, where the benefits of our proposed estimator are highlighted. In this section, we also {  detail how estimators of $\Sigma$ may be employed to improve predictions in classification problems. The method is applied to four datasets to demonstrate improvements in prediction accuracy}. All proofs are presented in the Supplement.

\section{Proposed batch-means estimator}
\label{sec:BM_estimator}

\subsection{General batch-means estimator}
\label{online:batching:ebs}

Batch-means estimators and its variants are critical components of output analysis methods in steady-state simulation. A general batch-means estimator can be set up in the following way. For iteration size $n$, the SGD iterates (after some user-chosen warm-up) are divided into $K$ batches with batch-sizes $b_{n,1}, \ldots, b_{n,K}$. Define $\tau_0 = 0$ and let $\tau_k = \sum_{j = 1}^k b_{n,j}$ for $k = 1, 2, \ldots, K$, denote the ending index for the $k$th batch. Then the batches are:
\begin{align*}
\underbrace{\{\theta_1, \ldots, \theta_{\tau_1}\}}_{1^{\text{st}}\text{ batch}}, \underbrace{\{\theta_{\tau_1 + 1},\ldots,\theta_{\tau_2}\}}_{2^{\text{nd}}\text{ batch}},\ldots, \underbrace{\{\theta_{\tau_{K-1}+1},\ldots,\theta_{\tau_K}\}}_{K^{\text{th}}\text{ batch}}.
\end{align*}
Let $\bar \theta_k = {b}_{n,k}^{-1}\sum_{i = \tau_{k-1}+1}^{\tau_{k}} \theta_i$ denote the mean vector of the $k^{\text{th}}$ batch. A general batch-means estimator is
\begin{equation}
\label{eq:bm-general}
\hat{\Sigma}_{\text{gen}} = \dfrac{1}{K} \sum_{k=1}^{K}b_{n,k}  \left(\bar{\theta}_k - \hat{\theta}_n \right)\left(\bar{\theta}_k - \hat{\theta}_n \right)^{\top}\,.
\end{equation}

Batch-means estimators of limiting covariances are commonplace in steady-state simulation { \citep{alexopoulos2004batch,chen1987multivariate,chien1997large,glynn1991estimating,munoz1997batch,song1995optimal} } and  MCMC { \citep{chakraborty2022estimating,liu2018weighted,mcmc:stopping,flagel2010}}. Their performance is critically dependent on the batching structure; this choice is process dependent and much work has gone into their study for ergodic and stationary Markov chains \citep{damerdji1995mean,liu2022batch}.

In the context of SGD, batch-means estimators were recently adopted in the sequence of works by \cite{chen2020aos,dong2021,chen2021jasa}. For $\eta_i = i^{-\alpha}$,  $\alpha \in (1/2, 1)$, the batch-size chosen by \cite{chen2021jasa} is
\begin{equation}
\label{eq:chen_batch}
b_{n,k} \propto k^{ \frac{1+\alpha}{1 - \alpha}} \,.
\end{equation}
The above choice is motivated by the following argument: if $b_{n,k}$ is reasonably large, the batch-mean vector is approximately normally distributed. Using \citet[][Equation 15]{chen2020aos}, for large $j$ and $k$ ($>j$), the strength of correlation between $\theta_j$ and $\theta_k$ is
\begin{align*}\label{corr:strength}
\prod_{i=j}^{k-1}\|I_d-\eta_{i+1}A\| \leq \exp\left( -\lambda_{\min} (A) \sum_{i=j}^{k-1}\eta_{i+1}\right),\numberthis
\end{align*}
where $\lambda_{\min} (A)$ is the smallest eigenvalue of $A$. Consequently, if $\sum_{i=j}^{k-1}\eta_{i+1}$ is sufficiently large, the $j^{\text{th}}$ and $k^{\text{th}}$ iterates are approximately uncorrelated. 
Therefore, for large $b_{n,k}$'s, batch-mean vectors are approximately independent and normally distributed. The batch-size in \eqref{eq:chen_batch} is such that $\sum_{i = j}^{k-1} \eta_{i+1}$ is sufficiently large. This reasoning ignores the dangers of choosing large batch-sizes. For any given iteration length, larger batch-sizes implies smaller number of batches leading to high variance and/or singular estimators of $\Sigma$. Consequently, the quality of inference is challenged and multivariate inference becomes difficult.

\subsection{Proposed batching strategy}
\label{sec:prop_batching}

Under an equal batch-size strategy, $b_{n,k} = b_n$ for all $k$; the number of batches is $a_n := K = \lfloor n/b_n \rfloor$. With this choice, the estimator in \eqref{eq:bm-general} simplifies to
\begin{align*}
\label{online:var:estimator}
\hat{\Sigma}_{b_n}=a_n^{-1}\sum_{k=1}^{a_n}b_n(\bar \theta_k-\hat \theta_n)(\bar \theta_k-\hat \theta_n)^\top=&~ \cfrac{b_n}{a_n}\displaystyle \sum\limits_{k=1}^{a_n}\bar \theta_k\bar \theta_k^\top-b_n\hat \theta_n\hat \theta_n^\top.\numberthis
\end{align*}

\begin{figure}[]
\centering
\includegraphics[width = 2.7in]{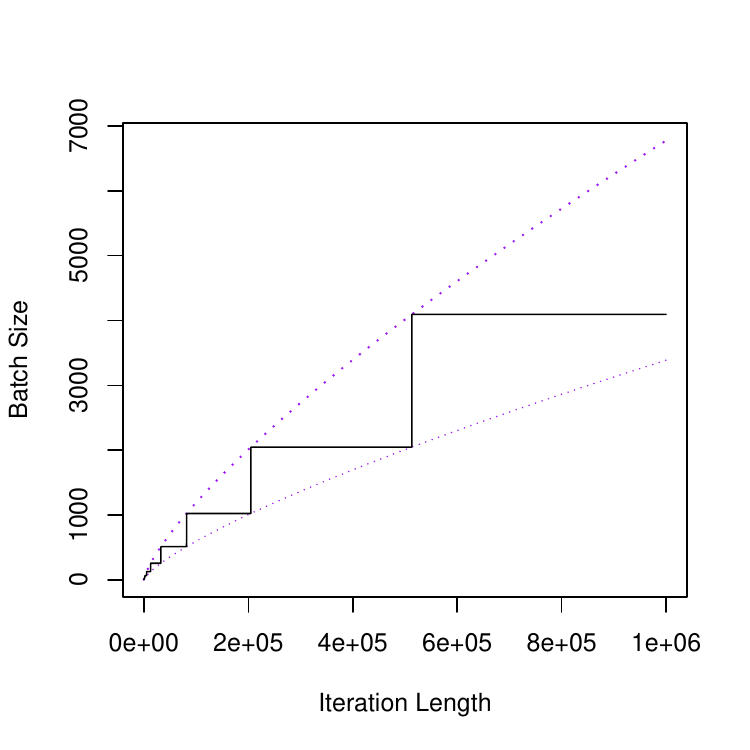}
\includegraphics[width = 2.7in]{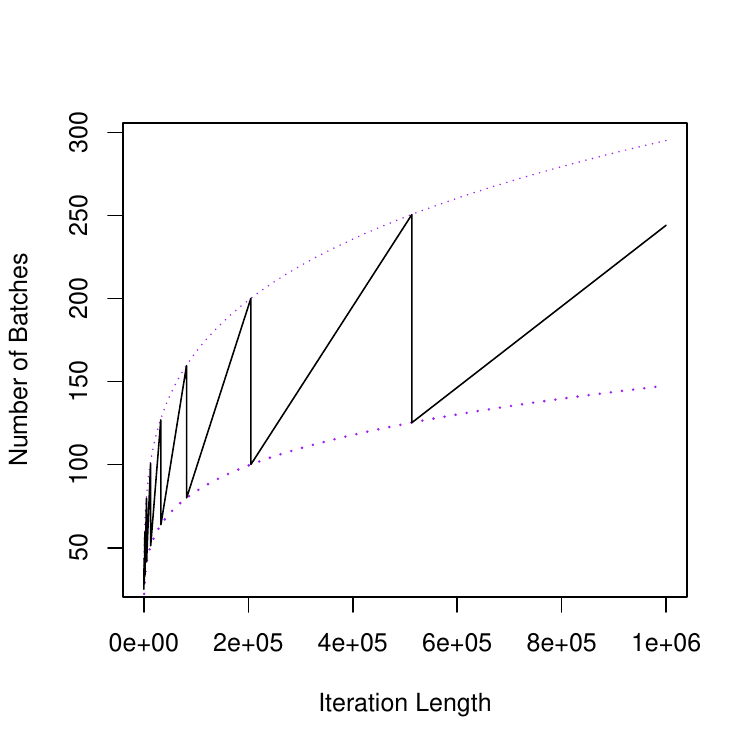}
\caption{Pictorial demonstration of the proposed batching structure, as a function of $n$; settings are chosen in accordance with the simulations in Section~\ref{sec:sims}. Purple dotted lines are polynomial rates that bound the batch-size (left) and number of batches (right).}
\label{fig:batch_structure}
\end{figure}

Choosing $b_n \propto \lfloor n^{\beta} \rfloor$ for some $\beta \in (0,1)$ seems natural, and is often considered in stochastic simulation. { We consider memory-efficient batch-sizes that are powers of two, and still grow polynomially. That is,} for some $c > 0$ and current iterate $n$, {  we} consider batch-sizes of the following form:
\begin{equation}
\label{eq:2k_batch_size}
b_n^* = \min\{ 2^{\gamma} :  cn^{\beta} \leq  2^{\gamma} \text{ for } \gamma \in \mathbb{N} \}\,.
\end{equation} 
That is, $b_n^*$ is the smallest power of 2 that is bounded below by $cn^{\beta}$. Naturally, $cn^{\beta} \leq b_n^* \leq 2cn^{\beta}$, and $a_n^* := n/b_n^*$ for any given $n$ is bounded like $n^{1 - \beta}/(2c) \leq a_n^* \leq n^{1 - \beta}/c$. A similar batch-size strategy was hinted at in \cite{gong2016practical}. A pictorial demonstration of the batching strategy is in Figure~\ref{fig:batch_structure}, for the settings discussed  in Section~\ref{sec:sims}. Naturally, as $n \to \infty$, both $b_n^*$ and $a_n^*$ tend to $\infty$. The proposed batching structure reduces storage costs to only $a_n$ batch-mean vectors at any given stage; $a_n = O(n^{1- \beta})$ dramatically smaller than $n$. Further, for the iterate of $n$ when batch-size changes, new batches are just made by averaging over adjacent batch-mean vectors; at these moments the number of batches gets halved. { Our theoretical results hold for a general of equal batch-sizes and in our simulations we implement both $b_n^*$ and $b_n = \lfloor cn^{\beta}\rfloor$.}
	 
\section{Main results}\label{sec:main:results}
	
Consistency of $\hat{\Sigma}_{b_n}$ along with the asymptotic normality result of \cite{pj1992} in \eqref{eq:pj_normal}, allows for large-sample inferential procedures, similar to traditional maximum likelihood estimation. For this task, we make assumptions that ensure both the asymptotic normality in \eqref{eq:pj_normal} and consistency of the covariance estimator.
	
\subsection{Notations and assumptions}

For a vector $x \in \mathbb{R}^d$, let $\| x\|$ denote the Euclidean norm and for a matrix $A$, let  $\|A\|$ denote its matrix norm. All norms are equivalent in a finite-dimensional Euclidean space, so in the following discussion, we can replace the matrix norm with any other norm.

%
\begin{itemize}
\item[(A1)] (On $F$). Let the objective function $F$ be such that the following hold:
\begin{enumerate}[(i)]
\item $F(\theta)$ is continuously differentiable and strongly convex with parameter $M > 0$. That is, for any $\theta_1$ and $\theta_2$,
\begin{equation}
\label{eq:assum_convex}
F(\theta_2)\geq F(\theta_1)+ \nabla F(\theta_1) ^{\top} (\theta_2-\theta_1)  + \frac{M}{2}\|\theta_2 - \theta_1\|^2.		
\end{equation}
\item The gradient vector $\nabla F(\theta)$ is Lipschitz continuous with constant $L_F$, that is, for any $\theta_1$ and $\theta_2$, 
\begin{equation}
\label{eq:assum_lipsch}
\|\nabla F(\theta_1)-\nabla F(\theta_2)\|\leq L_F\|\theta_1-\theta_2\|\,.			
\end{equation}
\item The Hessian of $F$ at $\theta^*$, $A=\nabla^2F(\theta^*)$ exists, and there exists $L_1$ such that $$\|A(\theta-\theta^*)-\nabla F(\theta)\|\leq L_1\| \theta - \theta^*\|^2.$$
\end{enumerate}
\end{itemize}
Assumption~(A1) is important for the convergence and  asymptotic normality of $\hat{\theta}_n$ \citep[see][]{pj1992,bach2011,rakhin}. The strong convexity of $F$ implies that $\lambda_{\min}(A)\geq M$, which is an important condition for parameter estimation (see \citealt{chen2020aos,chen2021jasa,dong2021}). Further, this will be a key ingredient for proving consistency of the batch-means estimator of $\Sigma$.
\begin{itemize}
\item[(A2)]\label{ass:a2} (On $f$ and $\zeta_i$). Let $D_i=\theta_i-\theta^*$, $\xi_i=\nabla F(\theta_{i-1})-\nabla f(\theta_{i-1},\zeta_i)$, and $\mathbb{E}_i(\cdot)$ denotes the conditional expectation $\mathbb{E}(\cdot | \zeta_i,\zeta_{i-1},\ldots,\zeta_1)$, then the following hold:
\begin{enumerate}[(i)]
\item The function $f(\theta,\zeta)$ is continuously differentiable in $\theta$ for any $\zeta$ and $\|\nabla f(\theta,\zeta)\|$ is uniformly integrable for any $\theta$ (so that $\mathbb{E}_{i-1}(\xi_i)=0$).
			
\item The conditional covariance of $\xi_i$ has an expansion around $\theta = \theta^*$. That is, $\mathbb{E}_{i-1}[\xi_i\xi_i^\top] = S+\mathcal{H}(D_{i-1})$, and there exist constants $\sigma_1$ and $\sigma_2$ such that for any $D\in\mathbb{R}^d$,
\begin{align*}
\|\mathcal{H}(D)\|\leq \sigma_1\|D\|+\sigma_2\|D\|^2\text{ and }
|\tr(\mathcal{H}(D))|\leq \sigma_1\|D\|+\sigma_2\|D\|^2.
\end{align*}
\item There exists constants $\sigma_3$ and $\sigma_4$ such that the fourth conditional moment of $\xi_i$ is bounded, i.e., $\mathbb{E}_{i-1}\|\xi_i\|^4\leq\sigma_3+\sigma_4\|D_{i-1}\|^4$.
\end{enumerate}
\end{itemize}
Assumption~(A2)(i) allows $\mathbb{E}_{\zeta\sim\Pi} \left[\nabla f(\theta,\zeta)\right] = \nabla F(\theta)$, which implies that the sequence $\{\xi_i\}$ is a martingale difference process. These assumptions are standard \citep[see][]{chen2020aos,chen2021jasa} and ensure the regularity of the noisy gradients. 

Our next set of conditions are on the learning rate and the choice of equal batch-size. Our results are general for any choice of batch-size satisfying the condition below.
\begin{itemize}
\item[(A3)] (On $\eta_i$, $b_n$) The following hold:
\begin{enumerate}[(i)]
\item The learning rate is $\eta_i=\eta i^{-\alpha}$ with $\alpha\in(0.5,1), i = 1, 2, 3, \ldots~$.
\item $b_n$ is size of the batch such that $b_n n^{-\alpha}\to\infty$ and $b_nn^{-1}\to0$ as $n\to\infty$. 
\end{enumerate}
\end{itemize}
In Assumption~(A3)(i), the learning rate is that of \cite{pj1992}, ensuring asymptotic normality. Assumption(A3)(ii) is the only additional condition added to this statistical inference setup that is specific to our choice of { equal} batch-size. Our chosen $b_n = c n^\beta$  satisfies this condition for  $\beta\in(\alpha,1)$ (and thus $b_n^*$ satisfies this condition as well). As a consequence of Assumption~(A3)(i),
\begin{align}\label{batch_sum_of_learning_rate}
\sum_{i = \tau_{k-1} + 1}^{\tau_k}\eta_i = \sum_{i=\tau_{k-1} + 1}^{\tau_k} \eta i^{-\alpha} > \eta b_n{\tau_k}^{-\alpha} > \eta b_nn^{-\alpha}=:N.
\end{align}
Using Assumption~(A3)(ii), $N\to\infty$ as $n\to\infty$. This guarantees that the batch-size is larger than the persistent correlation in the SGD iterates. That is, using \eqref{corr:strength}, this ensures fast decay of correlation between batches, which is a critical step in proving consistency of the batch-means estimator. 

In the following discussion, for sequences of positive numbers $p_n$ and $q_n$, denote
\begin{itemize}
\item  $ p_n \gtrsim q_n$ if for some $c > 0, c p_n \geq q_n$ for all $n$ large enough,
\item 	 $ p_n \lesssim  q_n$ if $ q_n \gtrsim p_n$, and
\item $p_n \asymp  q_n$ if $ p_n\gtrsim q_n$ and $ p_n\lesssim q_n$.
\end{itemize}
For simplicity, we define the following constant, 
\begin{align*}
C_d:=\max\left\{L_F,L_1,\sigma_1^{2/3},\sqrt{\sigma_2},\sqrt{\sigma_3}, \sigma_4^{1/4},\tr(S)\right\}.
\end{align*}
%
	
	
\subsection{Consistency of the estimator}

We now present our main result of consistency of the batch-means estimator {  for equal batch-sizes satisfying Assumption~(A3)}.
\begin{theorem}\label{thm:main}
Under the Assumptions (A1), (A2) and (A3), for  sufficiently large $n$
\begin{align*}			
~\mathbb{E}\big{\|}\hat{\Sigma}_{b_n}-\Sigma \big{\|} &\lesssim  ~C_d^{2}{n^{-\alpha/2}\,a_n^{-1/4}} + C_d^{3} n^{-\alpha} +C_da_n^{-1/2}+ C_db_n^{\alpha-1}+ C_db_n^{-1/2}n^{\alpha/2} \\
& \qquad + C_da_n^{-1}+ C_d^4{n}^{-2\alpha}{b_n}.
\end{align*}
\end{theorem}
\begin{proof}
Proof is available in the Supplement C.
\end{proof}
Under Assumption~(A3), the bound in Theorem~\ref{thm:main} goes to zero, yielding consistency of $\hat{\Sigma}_{b_n}$. \citet[][Theorem 4.3]{chen2020aos} and \citet[][Theorem 3.1]{chen2021jasa} provide similar bounds for different IBS batch-means estimators, with \cite{chen2021jasa} being an improvement over \cite{chen2020aos}. One key reason for explicitly writing a bound instead of merely mentioning convergence to zero, is that the bound allows for a reasonable choice for $b_n$. Substituting batch-sizes of the form $b_n = c n^{\beta}$, in Theorem \ref{thm:main}, 
\begin{align*}\label{n:beta:bound}
&~\mathbb{E}\big{\|}\hat{\Sigma}_{b_n}-\Sigma \big{\|}
\lesssim  ~n^{-\alpha/2+(\beta-1)/4}+ n^{(\beta-1)/2}+ n^{-\beta(1-\alpha)}+ n^{(\alpha - \beta)/2} + n^{\beta-1}+ {n}^{\beta -2\alpha}.\numberthis
\end{align*}
Obtaining a closed-form expression of an optimal $\beta$ from the right side of the above equation is challenging. A numerical solution is possible, but not interpretable. Instead, we note that by Assumption~(A3),
 $(\beta - 1)/2> \beta - 2\alpha$ and $-\alpha/2+(\beta-1)/4< (\beta - 1)/2$, so among the first, second, fifth, and sixth terms, the second term is dominating. Further, $-\beta(1-\alpha)< (\alpha - \beta)/2$, so among the third and fourth terms, the fourth term is dominating. Considering then, only the dominating terms, we have
\begin{align*}
\mathbb{E}\big{\|}\hat{\Sigma}_{b_n}-\Sigma \big{\|}
\lesssim  n^{(\beta - 1)/2} + n^{(\alpha - \beta)/2}.
\end{align*}
With this approximation, the optimal choice of $\beta$ is 
	$\beta^* = {(1+\alpha)}/{2}\,$.

\begin{remark}
The bounds we obtain are meaningful only for large $n$. For small $n$, it is challenging to obtain a meaningful expression of the optimal value of $\beta$ { in \eqref{n:beta:bound}}. Numerically, we observed that for sample size in the thousands and $\alpha = 0.51$, the optimal value of $\beta$ is near $0.66$. However, as $n$ increases, the optimal value of $\beta$ approaches $(1+\alpha)/2$. This agrees with the above mentioned bound. It is also important to remember that \eqref{n:beta:bound} is only a bound, optimizing which need not yield a true mean-square-optimal choice of batch-size.
\end{remark}

\begin{remark}
Consistency of $\hat{\Sigma}_{b_n}$ immediately allows the construction of Wald-like confidence regions (see Section~\ref{sec:regions}). To obtain a consistent estimator of $\Sigma$, the number of batches, $a_n$, must increase with $n$. Naturally, the batch-size also must be large to mimic the limiting \cite{pj1992} behavior. This yields a challenging trade-off. Our particular batch-size construction allows finite-sample adjustments for small batch sizes using the lugsail trick (see Section~\ref{sec:lugsail}). If the goal is only inference, and not the quantification of variance, 
 \cite{dong2021} {  used cancellation methods to construct valid confidence regions employing batch-means estimators with} fixed number of batches. { Further, such a method cannot be used for marginal friendly inference.}
\end{remark}
	
\begin{remark}{ 
With $\beta = (1 + \alpha)/2$, the computational complexity of calculating the batch-means estimator is similar to the IBS estimator at $\mathcal{O}(d^2 n^{(1 + \alpha)/2} + dn)$. On the other hand, an online implementation strategy for EBS estimators remains to be an open problem. Further, as we will discuss in Section~4, EBS estimators allow for bias-reduction strategies which yield significant benefits. Bias-reduction strategies in IBS estimators remain to be an open problem. }
\end{remark}

\section{Bias-reduced estimation}
\label{sec:lugsail}

Naturally, the mean-square bound in Theorem~\ref{thm:main} is contributed from both the bias and variance of the batch-means estimator.  \cite{vats:lugsail} proposed a lugsail batch-means estimator for stochastic simulation that can dramatically reduce bias in variance estimation. Our particular choice of equal batch-size allows an easy and effective implementation of the lugsail technique. Obtaining an exact expression of the bias of the batch-means estimator for a general SGD framework is an open problem. However, the following mean estimation model provides a motivation.

\begin{example}
\label{ex:mean_model} \normalfont  
	Consider for $i = 1, 2, \dots,n$, a mean estimation model $y_i = \theta^* + \epsilon_i$, 
	where $\theta^* \in \mathbb{R}$ and $\epsilon_i$ are independent mean-zero random error terms. For the squared error loss function $f(\theta, \zeta) = (y - \theta)^2/2$ for estimating $\theta^*$, the $i^{\text{th}}$ SGD iterate is
	\begin{align*}
		\theta_i=\theta_{i-1}+\eta_i (y_i-\theta_{i-1})\,,
	\end{align*}
	with $\eta_i = \eta i^{-{\alpha}}$. In the Supplement D.1, we show that the bias of $\hat{\Sigma}_{b_n}$ for this model is:
	\begin{align*}
		\Bias(\hat\Sigma_{b_n})\approx 
		\frac{-2C_1}{n}\sum_{1 \leq j < k \leq a_n} \sum_{p = \tau_{j-1}+1}^{\tau_j} \sum_{q = \tau_{k-1}+1}^{\tau_k} q^{-\alpha} (1 - q^{-\alpha})^{q - p}\,,
	\end{align*}
	where $C_1$ is a positive constant. The estimator of \cite{chen2021jasa} exhibits a similar negative bias expression. For large $n$, the bias may be insignificant, however, as Figure~\ref{fig:bias_mean_model} exhibits, even in this simple model, the finite-sample bias in the estimator of $\Sigma$ remains significant. 
	\begin{figure}[]
	\centering
	\includegraphics[width = 2.1in]{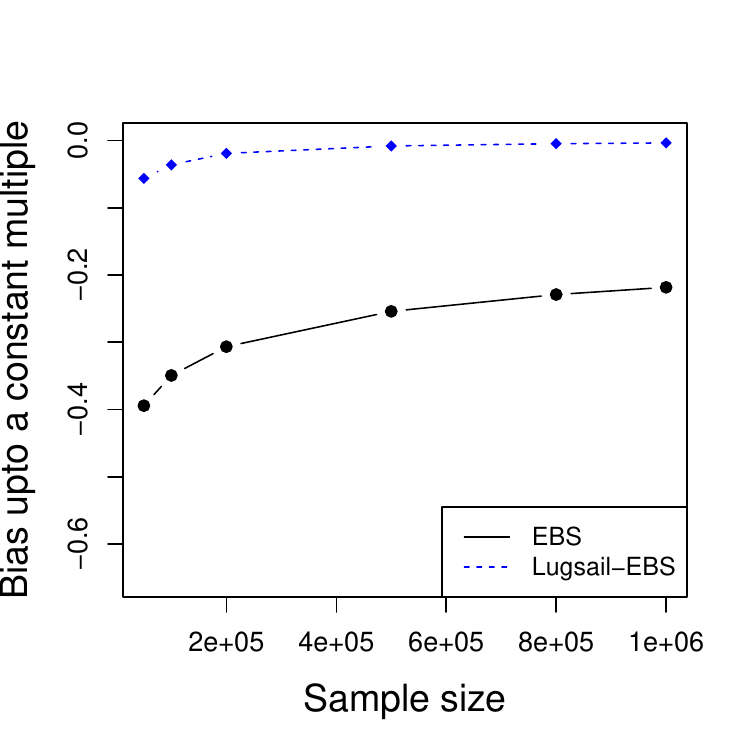}
	\includegraphics[width = 2.1in]{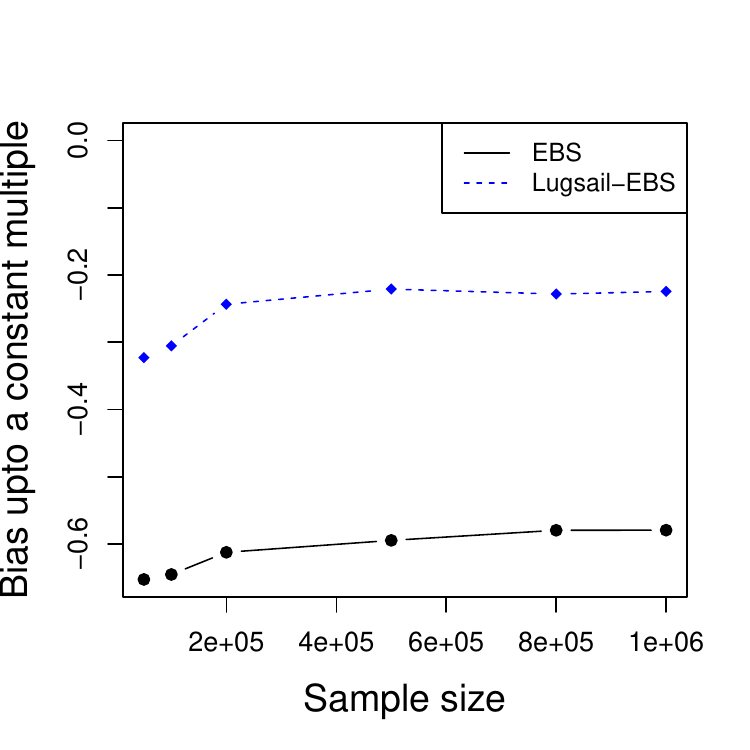}
	\caption{Mean estimation model: Bias of the EBS and Lugsail-EBS estimator for $\alpha=0.51$ (left) and $\alpha=0.75$ (right) against sample size.}
	\label{fig:bias_mean_model}
	\end{figure}	
\end{example}

More than the magnitude of bias, its direction is a larger concern. Variance estimation of any statistical estimator allows us to assess the uncertainty in the problem. Under-estimation of this variance leads to a false sense of security and inadequate { tests \citep[see][]{simon:1993}}. Obtaining bias-free estimators for such long-run variances is a critical and challenging problem in  operations research, stochastic simulation, econometrics, and MCMC. A wide range of solutions have been attempted \citep{kief:vogel:2002b,kief:vogel:2005,liu2018weighted,politis1995bias} to reduce the bias of variance estimators. 


{ 
By studying linear combinations of lag-windows in spectral variance estimators, \cite{liu2018weighted,vats:lugsail} develop a family of variance estimators, called \textit{lugsail} estimators, that are essentially  obtained by a carefully chosen linear combination of variance estimators.  \cite{liu2018weighted} define a batch-means version of this estimator called the weighted batch-means estimator, that seek to combine batch-means estimators obtained through various batch-sizes, using an appropriate weighting strategy. Consider a weighting function, called a \textit{lag window}, $w_n(\cdot)$ such that (i) $w_n(\cdot)$ is an even function on $\mathbb{Z}$, (ii) $w_n(0) = 1$ for all $n$, and (iii) $w_n(m) = 0$ for all $m \geq b_n$. \cite{anderson2011statistical} provides a comprehensive list of lag windows used in stochastic simulation and time series. We will employ the flat-top lag window of \cite{politis1995bias}:
\begin{equation}
\label{eq:flat_top}
    w_n(m) = \mathbb{I}\left( |m| \leq \dfrac{b_n}{2} \right) + 2 \left(1 - \dfrac{|m|}{b_n} \right) \mathbb{I}\left( \dfrac{b_n}{2} < |m| \leq b_n \right)\,.
\end{equation}
Let $\Delta_2 w_n(m) = w_n(m-1) - 2 w_n(m) + w_n(m+1)$ denote the second order differencing of $w_n$ at $m$. Consider multiple batch-sizes $m = 1, 2, \dots, b_n$, so that the corresponding number of batches are $\tilde{a}_m := \lfloor n/m \rfloor$. For each batch size $m$, let $\bar{\theta}_{m,k} = m^{-1} \sum_{t = 1}^{m} \theta_{km + t}$ for $k = 0, 1, 2, \dots, \tilde{a}_m$, denote the $k^{\text{th}}$ batch mean vector. The weighted batch-means estimator of \cite{liu2018weighted} is defined as
\begin{equation}
\label{eq:wbm}
    \hat{\Sigma}_{\text{wBM}} = \sum_{m=1}^{b_n}  \dfrac{m^2 \Delta_2 w_n(m)}{a_m} \sum_{k=0}^{a_m - 1}   \left( \bar{\theta}_{m,k} - \hat{\theta}_n \right)  \left(\bar{\theta}_{m,k} - \hat{\theta}_n  \right)^{\top} \,.
\end{equation}
For general lag windows, $\hat{\Sigma}_{\text{wBM}}$ can be expensive to compute due to the double summation in \eqref{eq:wbm}. However, employing piecewise linear lag windows like the flat-top lag window in \eqref{eq:flat_top}, yields $\Delta_2 w_n(m) = 0$ everywhere except $m = b_n/2, b_n$, making the estimator in \eqref{eq:wbm} computationally viable. 

\cite{liu2018weighted,vats:lugsail} discuss the bias-correction advantages of the weighted batch-means estimators. Employing \eqref{eq:flat_top} in \eqref{eq:wbm} and renaming $b_n \equiv 2b_n$, we obtain a bias-corrected batch-means estimator, compatible with equal batch-sizes, called the lugsail batch-means estimator.
}
Specifically, the lugsail batch-means estimator {  simplifies to}
	\begin{align}\label{sigma:lugsail}
		\hat \Sigma_{L,b_n} := 2\hat\Sigma_{2b_n} - \hat\Sigma_{b_n}.
	\end{align}
Our batching strategy, $b_n^*$, allows an easy implementation of the lugsail bias-correction strategy since a batch-means estimator of batch-size $2b_n^*$ can be obtained by { collapsing} adjacent batch-means vectors. Thus, the proposed bias-correction does not increase memory costs. We also note that since the lugsail lag-windows rely on equal batch-sizes, such lugsail corrections are not directly possible for the IBS estimator. We call the estimator in \eqref{sigma:lugsail} the lugsail-EBS estimator.

{ 
Define
\begin{equation}
    \label{eq:lugsail_remainder}
\hat{R}_{b_n} := \dfrac{b_n}{a_n}\sum_{j=1}^{a_n/2} \left[ \left(\bar \theta_{2j-1}-\hat \theta_n \right)  \left(\bar \theta_{2j}-\hat \theta_n \right)^\top + \left(\bar \theta_{2j}-\hat \theta_n \right)  \left(\bar \theta_{2j-1}-\hat \theta_n \right )^\top \right]\,.
\end{equation}
Then, $\hat{R}_{b_n}$ summarizes the covariance between adjacent batches. In Supplement~D.2  we show that
\begin{equation}
\label{eq:lugsail_decomp}
\hat{\Sigma}_{L,b_n} = \hat{\Sigma}_{b_n} + 2\hat{R}_{b_n}\,.
\end{equation}
When batch-sizes are not large enough, adjacent batch means will be expected to be positively correlated so that (the diagonals of) $\hat{R}_{b_n} > 0$, thereby adjusting some of  systematic under-estimation happening due to a small batch-size.
}
For the mean estimation model in Example~\ref{ex:mean_model},  Figure~\ref{fig:bias_mean_model}  presents the bias expression of both EBS and the lugsail-EBS estimators. Although the bias in the estimator depends on the correlation in the process (through $\alpha$ in this case), in both cases, the lugsail-EBS estimator presents significant bias reduction. As we will see in Section~\ref{sec:sims}, this correction proves to be critical for finite-sample inference.

The following results establish the consistency of the lugsail-EBS estimator, under the same conditions as required in Theorem~\ref{thm:main}.

{ 
\begin{prop} \label{new:lemma:lugsail:remainder}
Under Assumptions~(A1), (A2) and (A3), for sufficiently large $n$,
\begin{align*}
\mathbb{E}\|\hat{R}_{b_n}\| & \lesssim C_d^{1.25}  {n^{-\alpha/2}\,a_n^{-1/4}} + C_d^{2} n^{-\alpha/2}+ C_d a_n^{-1/2} + C_db_n^{\alpha-1} + C_db_n^{-1/2}n^{\alpha/2}\\ 
& \qquad + C_da_n^{-1}+ C_d^4{n}^{-2\alpha}{b_n}.
\end{align*}
\end{prop}
\begin{proof}
Proof is available in the Supplement D.3.
\end{proof}

{ An immediate consequence of the rate in Proposition~\ref{new:lemma:lugsail:remainder} is the following result that yields similar rates for the lugsail estimator as the original EBS batch-means estimator. Of course, finite-sample performance is affected by the absorbed constants, and we will see in Section~\ref{sec:sims} that the finite-sample performance of lugsail estimators is far improved.}
\begin{corollry}
Under Assumptions~(A1), (A2) and (A3), for sufficiently large $n$,
\begin{align*}
\mathbb{E}\big{\|}\hat \Sigma_{L,b_n}-\Sigma \big{\|}  \asymp \mathbb{E}\big{\|}\hat \Sigma_{b_n}-\Sigma \big{\|}. 
\end{align*}
\end{corollry}
\begin{proof}
Observe that the only different order term in Proposition \ref{new:lemma:lugsail:remainder} as compared to Theorem \ref{thm:main} is $n^{-\alpha/2}$.  Using Assumption (A3), for all $n$, $b_n^{1/2} n^{-\alpha}<  (b_n n^{-1})^{\alpha}<1$. Consequently, $b_n^{1/2} n^{-\alpha/2}n^{-\alpha/2}<1$, and $ n^{-\alpha/2}< b_n^{-1/2} n^{\alpha/2}$. Therefore, $n^{-\alpha/2}$ decays faster than $b_n^{-1/2} n^{\alpha/2}$. This illuminates that the rate for the bound on $\mathbb{E}\|\hat{R}_{b_n}\|$ is the same as that of $\mathbb{E}\|\hat \Sigma_{b_n}-\Sigma\|$. Using the triangle inequality, the result follows.
\end{proof}

\begin{remark}
Unlike standard EBS estimators that are guaranteed to be positive semi-definite, lugsail estimators may not retain this property, particularly  for small sample sizes. In such a case, users may tune $c$ so that $\hat{\Sigma}_{L,b_n}$ is positive-definite. In our simulations, $c$ was chosen appropriately so as to allow all estimators to be positive-definite.
\end{remark}

}



\section{Marginal and simultaneous inference}
	\label{sec:regions}

For problems where SGD is relevant, it is natural to ask what purpose will an estimator of $\Sigma$ serve? { Confidence ellipsoids provide little interpretation in high-dimensional settings, compared to hyper-rectangular regions that are more amenable to marginal-friendly interpretation.} \cite{chen2021jasa} use the diagonals of $\Sigma$ to construct uncorrected marginal confidence intervals for each of the $d$ parameters of interest. The problem of multiple-testing is omnipresent in this case, and corrections like Bonferroni can be crude. Moreover, the potentially complex dependence in $\Sigma$, via both $S$ and $A$ is completely ignored. { Here, we leverage the methods of \cite{marginal:friendly} to construct simultaneous marginal-friendly confidence regions that approximately retain the desired coverage probabilities.}

\begin{figure}[]
\centering
\includegraphics[width = 2.7in]{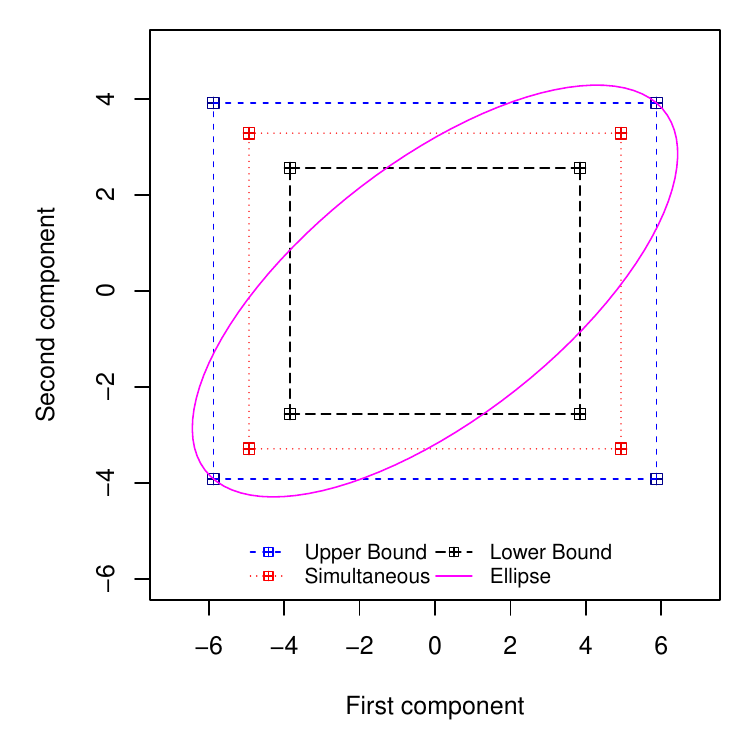}
\caption{Plot of $C_{lb}(z_{p/2})$ (black dashed) and $C_{ub}(z_{p/2d})$ (blue dashed) from a $90\%$ confidence region for a bivariate normal distribution with component variances 9 and 4. The red dashed line gives the corresponding $C({z^*})$.}
\label{fig:marginal:friendly:illustration}
\end{figure}

For joint inference for $\theta^*$, \eqref{eq:pj_normal} provides a $100(1 - p)\%$ confidence ellipsoid
	\begin{align}\label{eq:ellipsoid}
	  E_p  = \left\{\theta \in \mathbb{R}^d: (\hat\theta_n-\theta)^\top \hat\Sigma^{-1}_n(\hat\theta_n-\theta) \leq \chi^2_{d,1 - p} \right\},
	\end{align}
where $\chi^2_{d,s}$ denotes the $s^{\text{th}}$ quantile of a chi-squared distribution with $d$ degrees of freedom. Marginal interpretation of such an ellipsoid confidence region is difficult. Instead, one may study marginal confidence intervals.	Let $\hat\Sigma_n$ be any consistent estimator of $\Sigma$. Let $\hat\theta_n=(\hat\theta_{n1},\ldots,\hat\theta_{nd})^\top$, $\theta^*=(\theta^*_{1},\ldots,\theta^*_{d})^\top$ and $\hat \Sigma_n = (\hat\sigma_{ij})_{i,j = 1, \dots, p}$. Using \eqref{eq:pj_normal}, for $0<p<1$, an asymptotic $100(1-p)\%$ marginal confidence interval of $\theta^*_i$ is: 
\begin{equation}
	\label{eq:uncorrected}
	\hat\theta_{ni} \pm z_{1 - p/2}\sqrt{\hat\sigma_{ii}/n}\,,	\end{equation}
where $z_{s}$ denotes the $s^{\text{th}}$ quantile of $N(0,1)$. 
\cite{sgd:bootstrap}, \cite{chen2020aos}, \cite{chen2021jasa}, and \cite{dong2021} discuss both  uncorrected marginal confidence intervals and the ellipsoid joint confidence region. As discussed, both are inconducive for valid and interpretable joint inference. For the general Monte Carlo problem, \cite{marginal:friendly} suggest a remedy by using an appropriate hyper-rectangular confidence region, which we now describe. Using the $d$ uncorrected intervals in \eqref{eq:uncorrected},  an at-most $100(1-p)\%$ hyper-rectangular confidence region is 
\[
C_{lb}(z_{p/2})= \prod_{i=1}^d \left[\hat\theta_{ni}-z_{1 - p/2}\sqrt{\hat{\sigma}_{ii}/n},\,\hat\theta_{ni} + z_{1 - p/2}\sqrt{\hat{\sigma}_{ii}/n}\right]\,.
\]
Using a Bonferroni approach, an at least $100(1-p)\%$ hyper-rectangular confidence region is 
\[
C_{ub}(z_{p/2d})=\prod_{i=1}^d \left[\hat\theta_{ni}-z_{1 - p/2d}\sqrt{\hat{\sigma}_{ii}/n},\, \hat\theta_{ni} + z_{1 - p/2d} \sqrt{\hat{\sigma}_{ii}/n} \right]\,.
\]
Clearly, $C_{lb}(z_{p/2})\subseteq C_{ub}(z_{p/2d})$. 
\cite{marginal:friendly} used a quasi Monte-Carlo approach to find a $z^*$ with $z_{1 - p/2} < z^* < z_{1 - p/2d}$ to yield the hyper-rectangular confidence region 
\begin{align}\label{eq:marginal:friendly:inference}
C(z^*)= \prod_{i=1}^d  \left[\hat\theta_{ni} - z^*\sqrt{\hat{\sigma}_{ii}/n},\,\hat\theta_{ni} + z^*\sqrt{\hat{\sigma}_{ii}/n} \right]\,,
\end{align}
such that $C_{lb}(z_{p/2})\subseteq C(z^*)\subseteq C_{ub}(z_{p/2d})$ and $\mathbb{P}(\theta^*\in C(z^*))\approx 1-p$, under the assumption that $\hat{\theta}_n \approx N_p(\theta_*, \hat{\Sigma}_{n})$.  An illustration is given by Figure \ref{fig:marginal:friendly:illustration}.  The computation of  \eqref{eq:marginal:friendly:inference} is essentially a quick one-dimensional optimization problem solved by a bisection search over the interval $(z_{1 - p/2}, z_{1 - p/2d})$; see \cite{marginal:friendly} for details. In Section~\ref{sec:sims}, we present coverage properties of both the ellipsoidal region $E_p$ and the hyper-rectangular region $C(z^*)$.

\section{Numerical implementations}
\label{sec:sims}
	
\subsection{Setup} 
\label{sub:setup}
	
We implement our proposed EBS and lugsail-EBS estimators for two simulated models and a real data implementation, {  for both doubling batch-sizes $b_n^*$ and polynomial batch-sizes $b_n = \lfloor cn^{\beta}\rfloor$; we call these EBS and EBS-poly, respectively. Their lugsail versions are L-EBS and L-EBS-poly, repsectively.} We systematically keep the following settings for our EBS estimator: $c = 0.1$ so that the number of batches stays reasonably large ensuring that the estimators are positive-definite; $\beta = (1 + \alpha)/2$ as a reasonable value obtained from the mean-square bounds; $\alpha = 0.51$ to allow for reasonable exploration. For comparison we implement the IBS estimator of \cite{chen2021jasa}  with their suggested settings.  However, in the event that the IBS estimator is singular, we increase their number of batches to allow positive-definiteness of the IBS estimator.  { \cite{chen2021jasa} showed that their estimator was superior to the estimator of \cite{chen2020aos}, both theoretically and in simulations, so for brevity and clarity, we do not present comparisons with the estimator of \cite{chen2020aos}.}
	
When the true covariance matrix $\Sigma$ is available, we employ it as an oracle, and use it to calculate the relative Frobenius norm of an estimator $\hat \Sigma$: $\|\hat \Sigma-\Sigma\|_F/\|\Sigma\|_F$. Further, we employ $\Sigma$ in calculating { and comparing} the coverage probability of the confidence regions discussed in Section~\ref{sec:regions} { for different estimators}. Another important feature of confidence regions is its volume, particularly for the hyper-rectangular regions created in \eqref{eq:marginal:friendly:inference}. Thus, for each estimator, we also report
\[
\left(\dfrac{\text{Volume}(C(z^*))}{ \text{Volume} (E_p)} \right)^{1/p}\,\]
{ for which} a high value indicates an undesired increase in the volume of the { hyper-rectangular} confidence region. 

\subsection{Linear regression}
\label{linear:regression}
	
We simulate data according to the linear regression model, for $i = 1, 2, \dots, n$,  $y_i = x_i^T \beta^* + \epsilon_i$ where for some $d \times d$ positive-definite matrix $A$, $x_i \overset{\text{iid}}{\sim} N(0, A)$ and $\epsilon_i \overset{\text{iid}}{\sim} N(0,1)$, independent of $x_i$. We fix $\beta^* \in \mathbb{R}^d$ to be the $d$-vector of equidistant points on the grid $(0,1)$. Here $\zeta_i = (x_i, y_i)$. In order to implement ordinary least squares estimation of $\beta$, the loss function is
\[
f(\beta, \zeta_i) = \cfrac{\big(y_i - x_i^\top \beta\big)^2}{2}.
\]
Since the errors are { independent and identically distributed} (iid), the true $\Sigma$ is $A^{-1}$. We consider the three forms of $A$ used in \cite{chen2020aos}: (i) identity ($A = I_d$), (ii) Toeplitz, where element $A_{i,j} = \rho^{|i-j|}$, and (iii) equicorrelation, where element $A_{i,j} = \rho$ for $i \ne j$ and $1$ otherwise. Throughout, we set $\rho = 0.5$. We present results of all three settings of $A$ for $d = 5$ (here $\eta = 0.5$) and for brevity, only present the identity result for $d = 20$ (here $\eta = 1$).

Data of size $5\times 10^6$ was simulated with the first 1000 SGD iterates being discarded as burn-in. We start the SGD process from \(\bm{0}\) and study the statistical performance of various estimators of $\Sigma$ sequentially as a function of the data. Since the above is done for 1000 replications, for each $A$ and $d$, we present four key comparative plots: (i) the estimated relative Frobenius norm as a function of the sample size, (ii) the estimated ellipsoidal coverage probabilities, (iii) the estimated coverage probabilities of the hyper-rectangular confidence regions, and (iv) the ratio of the volumes of the hyper-rectangular regions to the ellipsoidal regions.   

\begin{figure}
    \includegraphics[width = 1.9in]{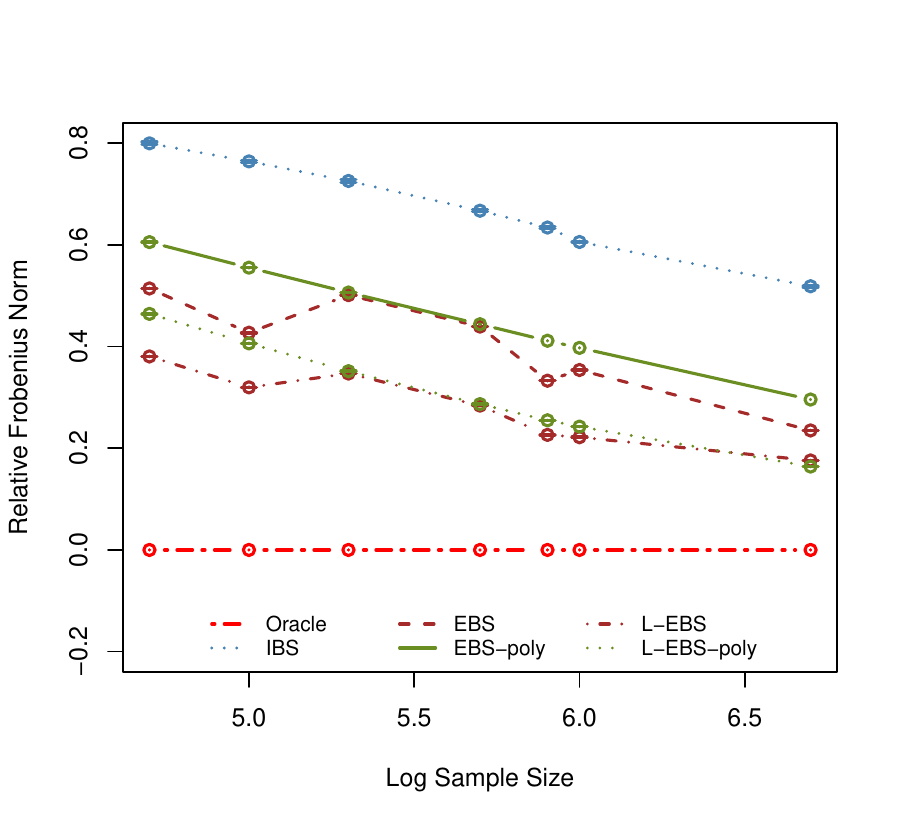}
    \includegraphics[width = 1.9in]{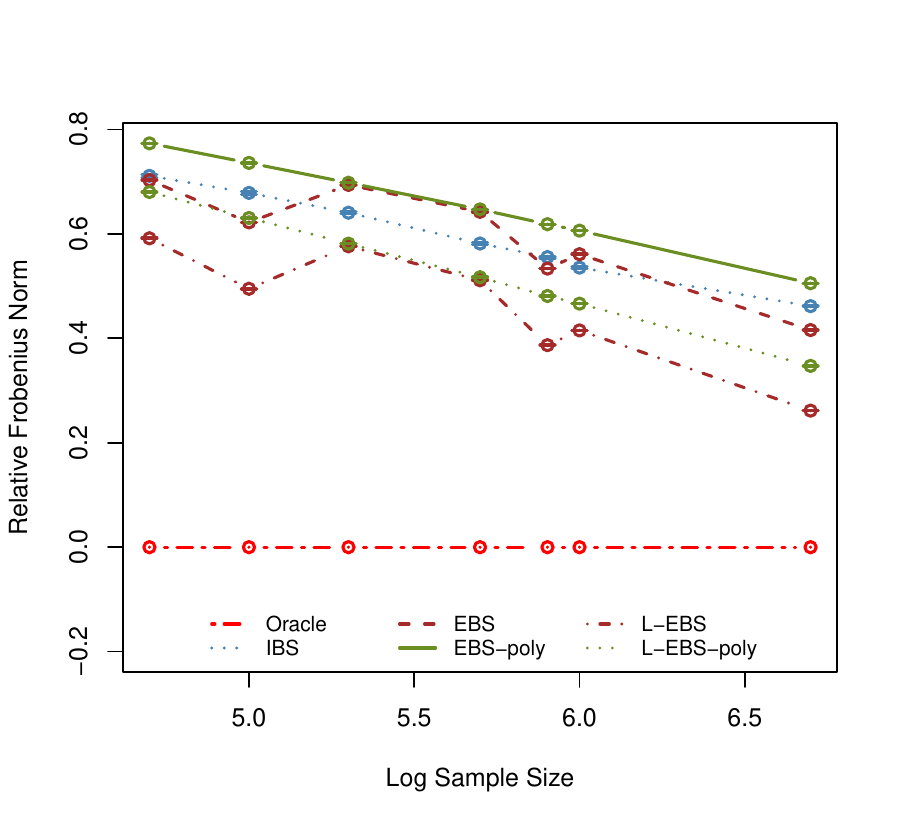}
    \includegraphics[width = 1.9in]{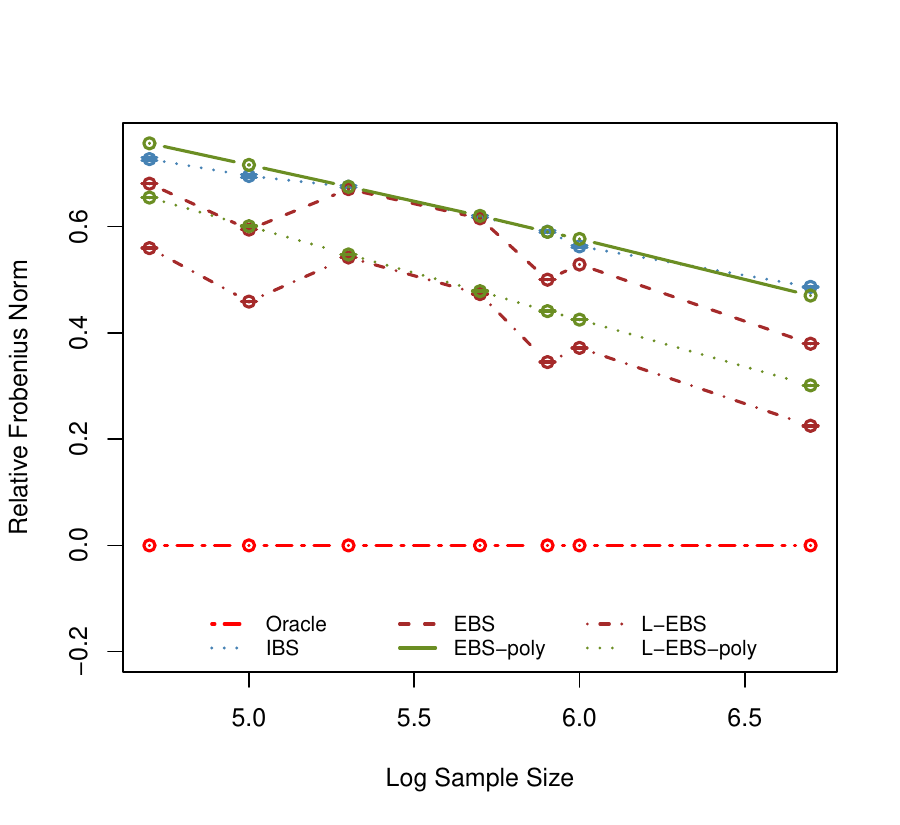}

    \includegraphics[width = 1.9in]{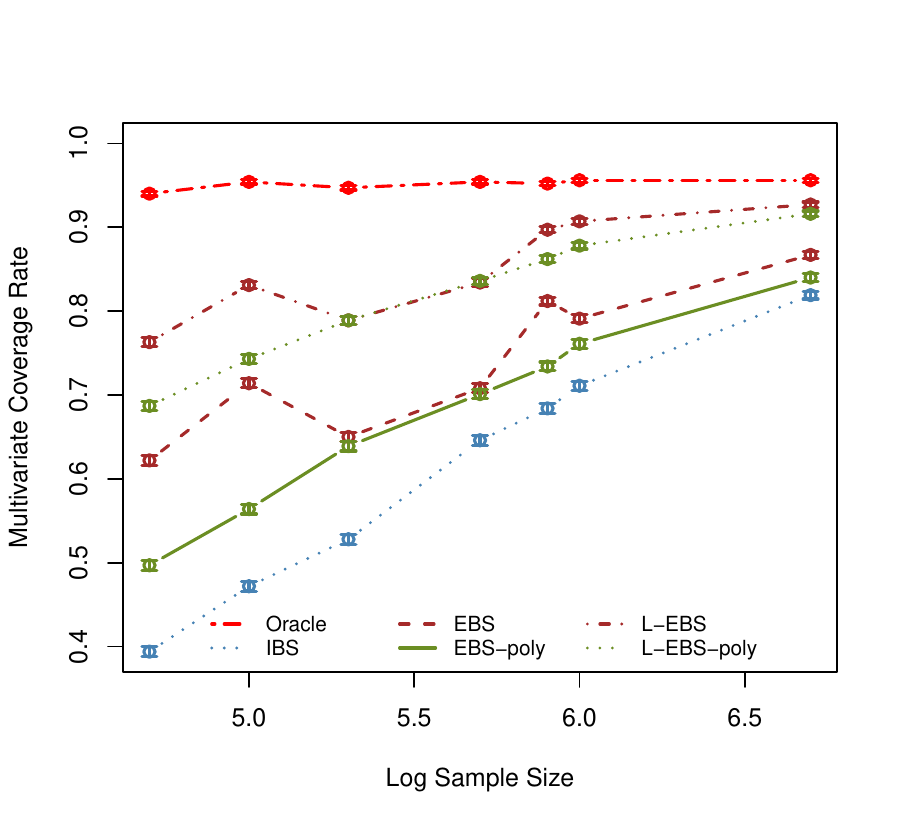}
    \includegraphics[width = 1.9in]{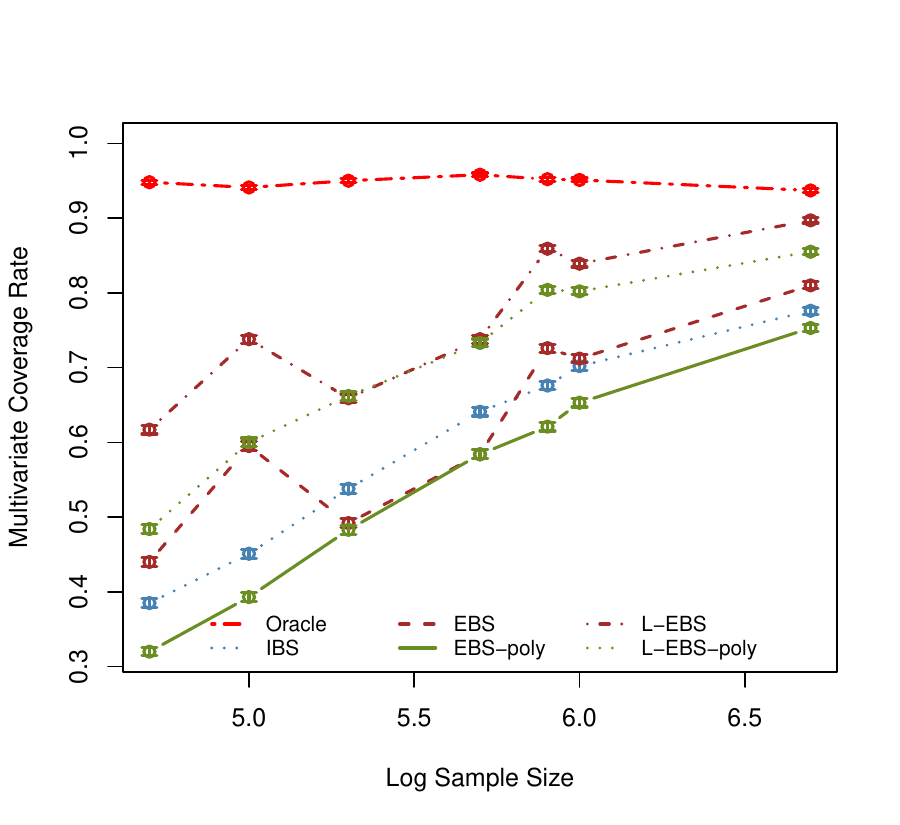}
    \includegraphics[width = 1.9in]{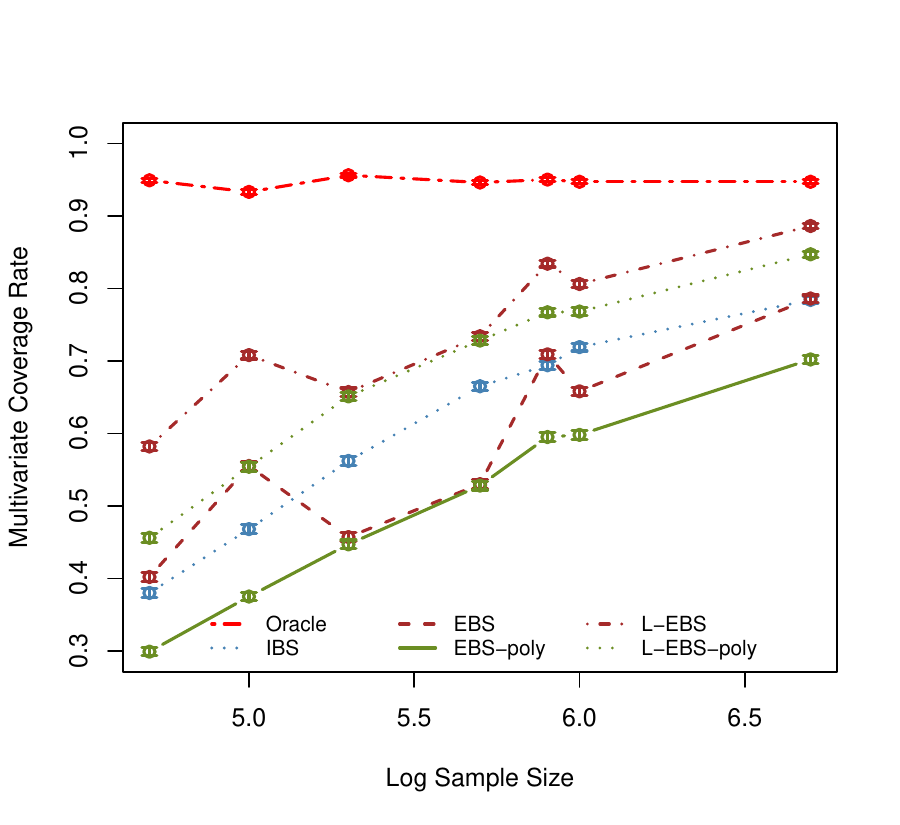}

    \includegraphics[width = 1.9in]{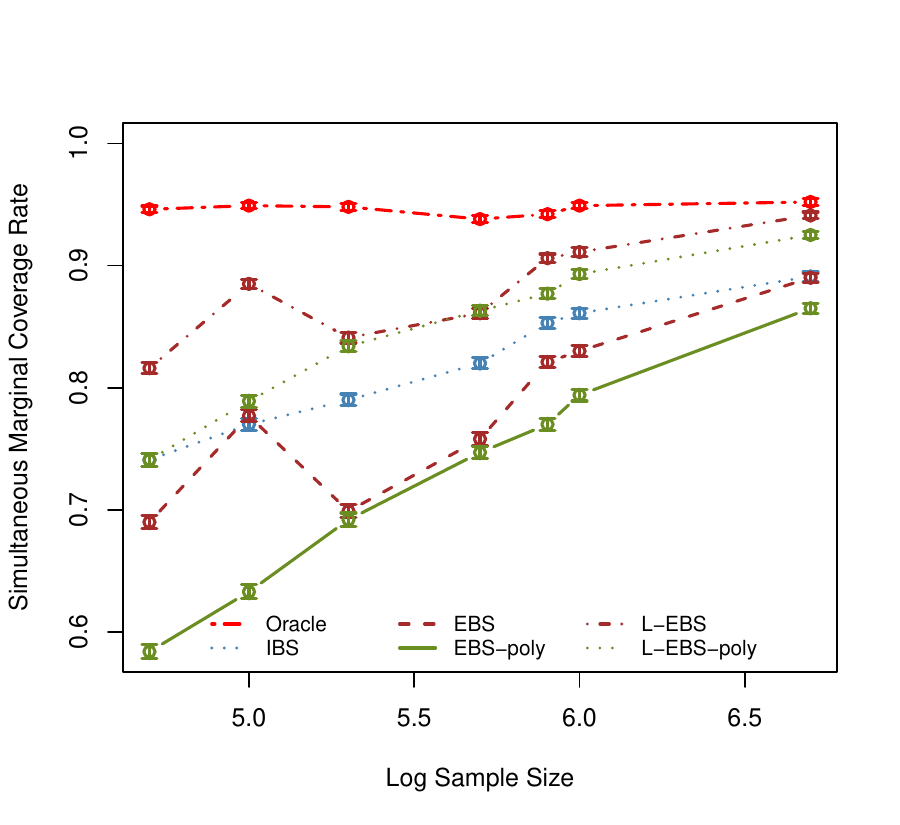}
    \includegraphics[width = 1.9in]{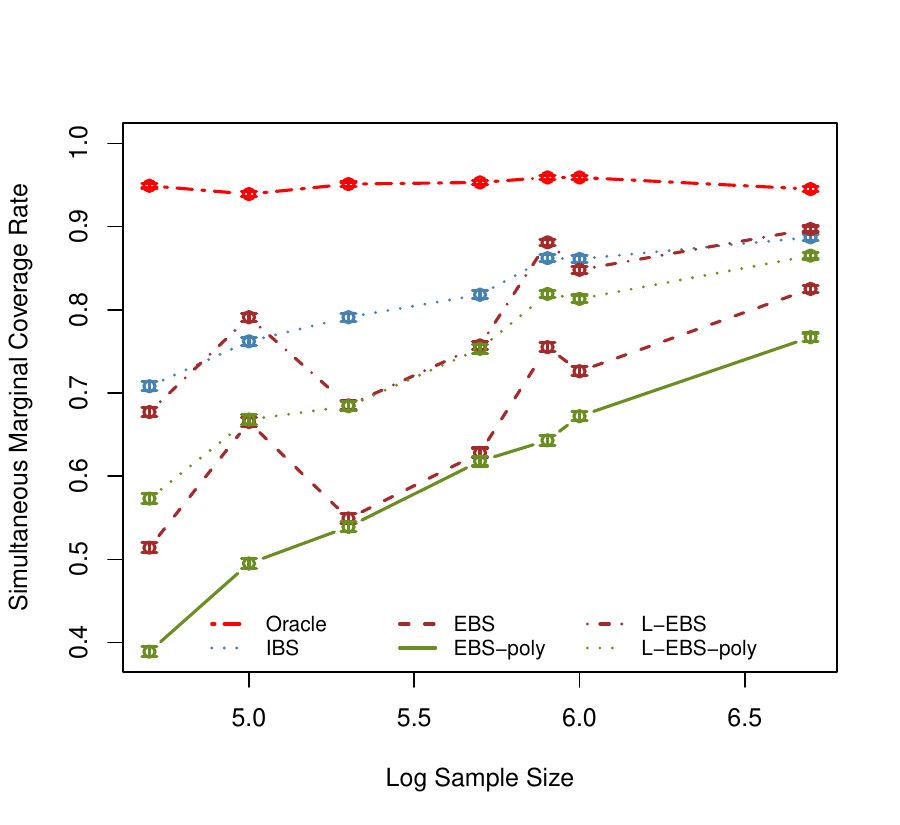}
    \includegraphics[width = 1.9in]{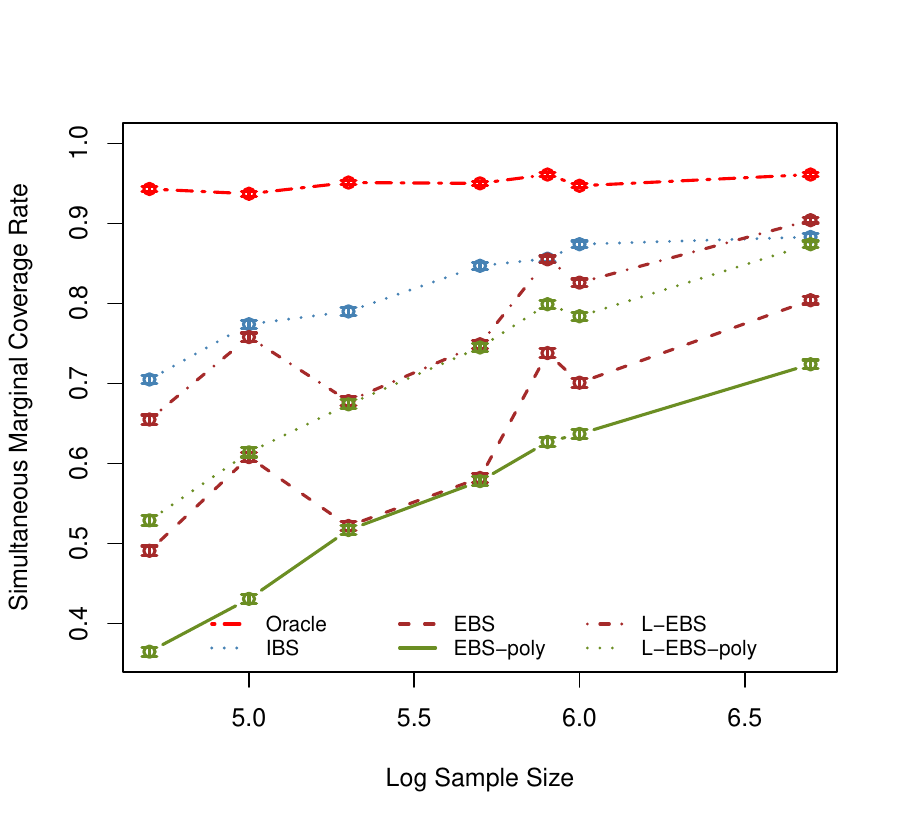}

    \includegraphics[width = 1.9in]{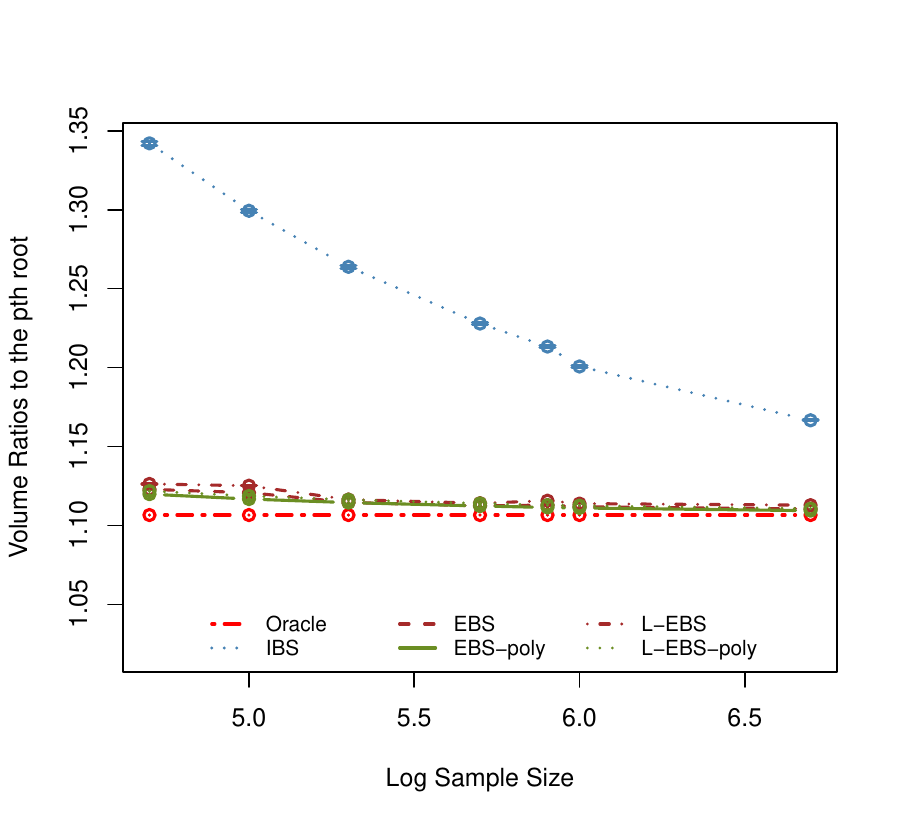}
    \includegraphics[width = 1.9in]{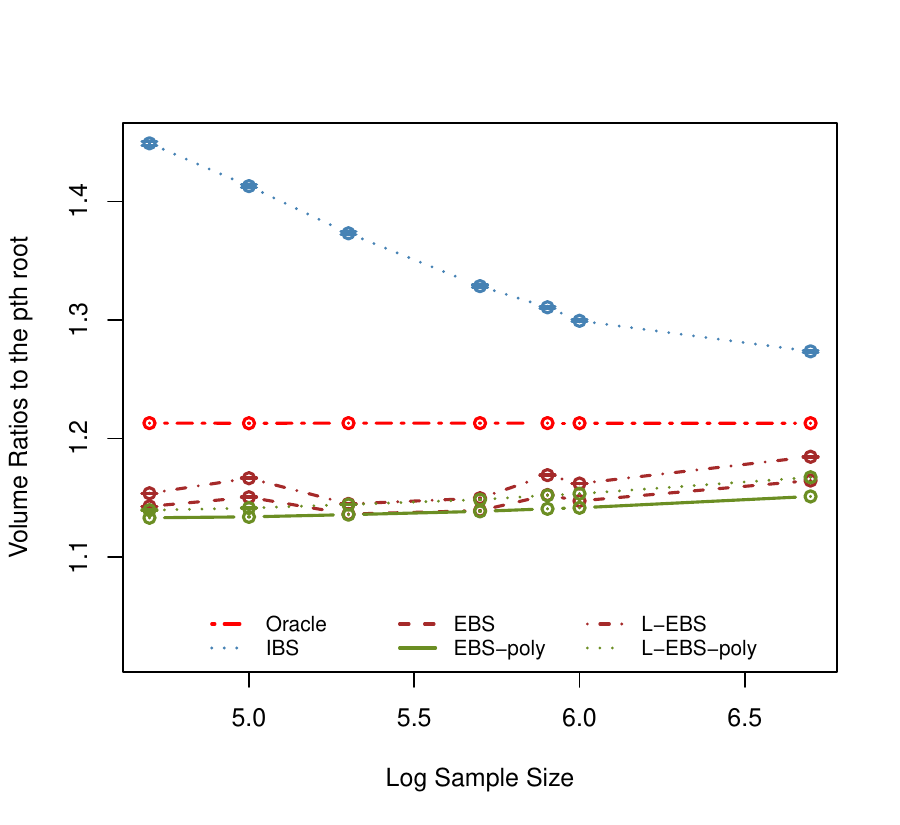}
    \includegraphics[width = 1.9in]{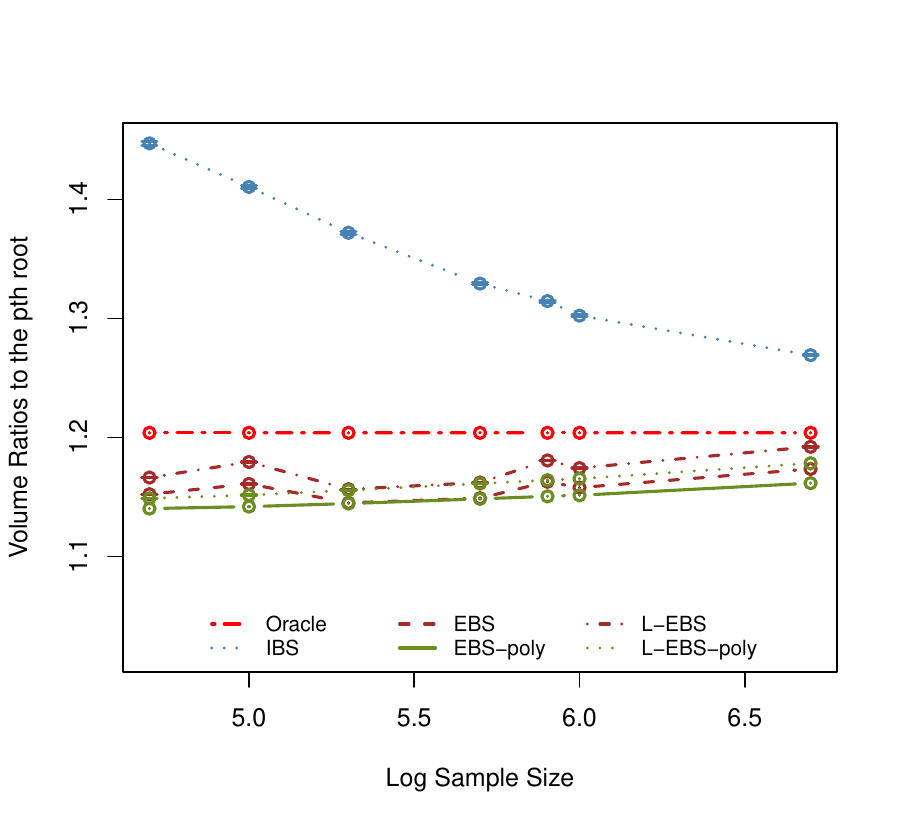}
    
\caption{Linear regression $d = 5$: Line plots equipped with error bars from 1000 replications. Left plots for $A$ being identity, middle plots for $A$ being Toeplitz, and right plots for $A$ being equicorrelation matrix. }
\label{fig:Lin_p5}
\end{figure}

\begin{figure}
    \centering
    \includegraphics[width = 2.1in]{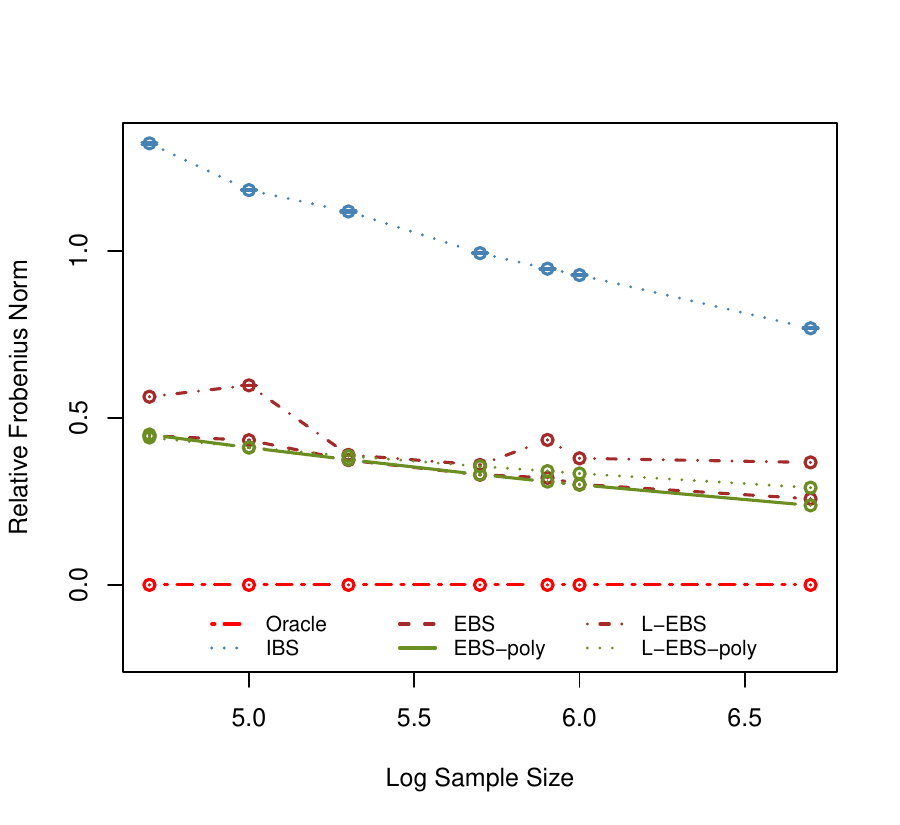}
    \includegraphics[width = 2.1in]{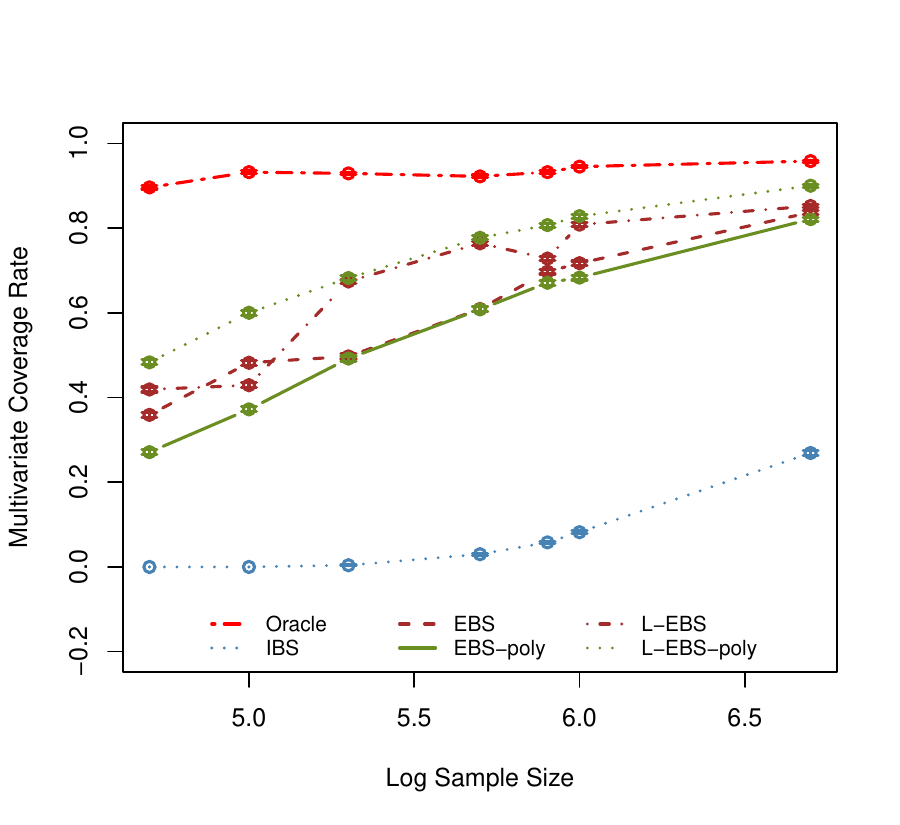}

    \includegraphics[width = 2.1in]{Lsimul_dim5.pdf}
    \includegraphics[width = 2.1in]{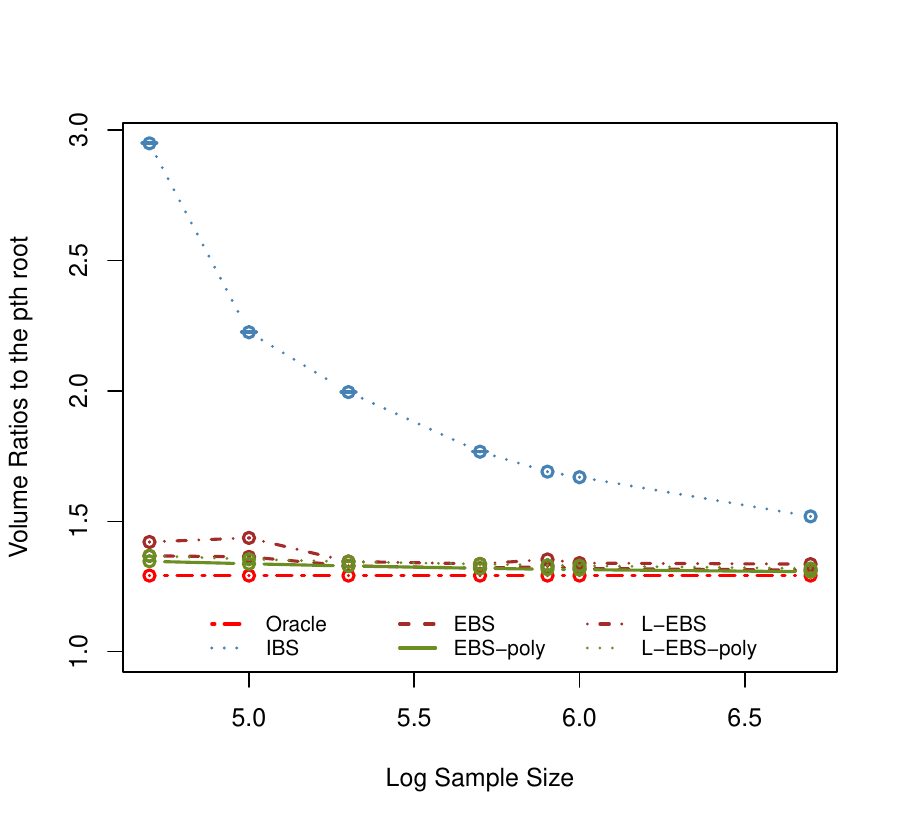}
    
    \caption{Linear regression $d = 20$ for identity $A$: Line plots equipped with error bars from 1000 replications}
    \label{fig:Lin_p20}
\end{figure}

Figure~\ref{fig:Lin_p5} presents the results for $d = 5$. Due to the nature of the {  doubling} batching technique {  $b_n^*$}, the performance is not monotonic as a function of the sample size for EBS; this is expected. { However, the polynomial batch-size in EBS demonstrates the expected monotonic behavior.} Further, for each of the three settings, the lugsail-EBS estimators outperforms in all measures with the EBS { estimators} being competitive with the IBS estimator, when not better. One metric where the IBS estimator suffers drastically, are the marginal confidence regions. As evident from Figure~\ref{fig:Lin_p5}, the coverage for the IBS estimator improves drastically when going from the elliptical regions to the hyper-rectangular regions. The bottom row of the plots indicate that this is entirely due to the drastic increase in the volume of region. The hyper-rectangular regions made by IBS are significantly larger that their ellipsoidal regions. This is likely due to an exaggerated correlation structure captured by the IBS. { All} the EBS estimators, and particularly the lugsail-EBS {  estimators} do not suffer from this. These problems are further exaggerated for $d = 20$ as evidenced in Figure~\ref{fig:Lin_p20}. The performance of both EBS and lugsail-EBS { estimators} remain essentially the same. We highlight that even when the relative Frobenius norm is large for the EBS estimators, the simultaneous marginal coverage probabilities are reasonable, with only little cost to the volume. 

\begin{figure}
    \centering
    \includegraphics[width = 2.5in]{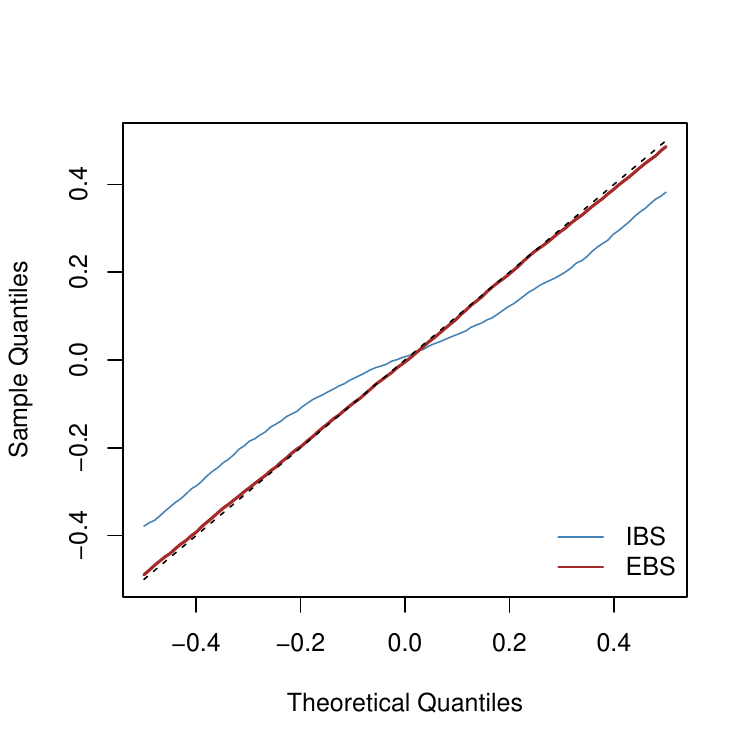}
    \includegraphics[width = 2.5in]{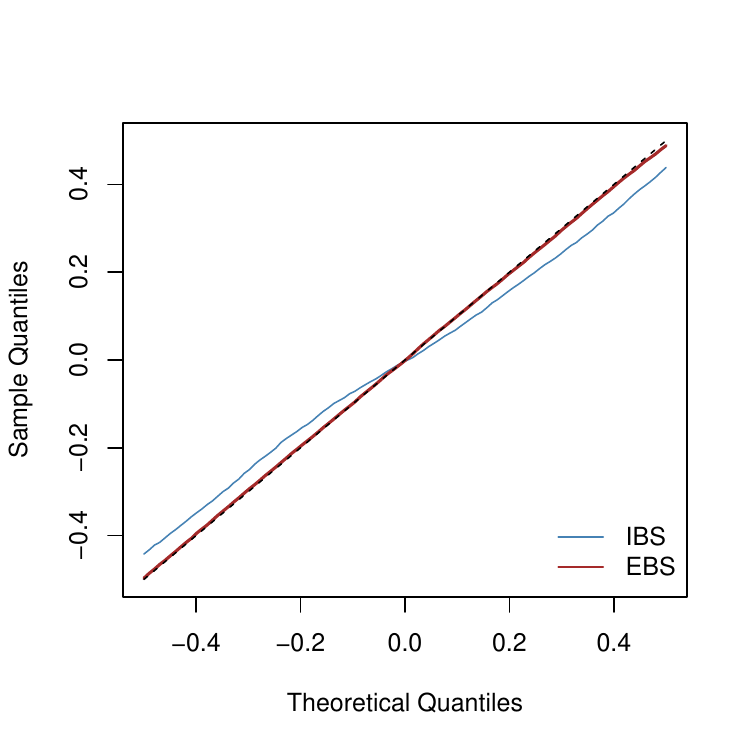}
    \caption{Linear regression for $d = 5$: QQ plot of all components of the batch-means vectors in the IBS and the EBS estimators, with black lines indicating theoretical quantiles. Left panel is corresponding to $n = 50000$ and right panel is corresponding to $n=10^6$. }
    \label{fig:batch:means}
    
\end{figure}

To understand why the EBS { estimators} perform significantly better than the IBS estimators, we take a closer look at the batching strategy { in $b_n^*$}. The general idea in batch-means estimators is that each batch-mean vector emulates the sample mean ASGD estimator; thus the empirical sample covariance of these batch-means is a reasonable estimator of $\Sigma$. For such a heuristic to hold, the batch-mean vector for each batch must be approximately normally distributed; $\sqrt{b_{n,k}}\bar{\theta}_k \approx N_{d}(\theta^*, I_d)$.  For $d = 5$ with $A$ being identity, if we accumulate all the components of all batch-mean vectors they should be normally distributed, and thus we may compare them with true Gaussian quantiles. In Figure~\ref{fig:batch:means}, we present a zoomed-in QQ plot of this for two different data sizes. Figure~\ref{fig:batch:means}  reveals significant deviation from normality for the IBS estimator, particularly for small sample situations. The EBS estimator, on the other hand, follows the theoretical Gaussian quantiles fairly well.

\subsection{LAD regression}
Assume a similar linear regression model, with non-Gaussian errors: $y_i = x_i^T \beta^* + \epsilon_i$ where $x_i \overset{\text{iid}}{\sim} N(0, A)$ and $\epsilon_i \overset{\text{iid}}{\sim} \text{DE}(0,1)$, here \( \text{DE}(\mu, \lambda)\) denotes the double exponential distribution with median parameter \(\mu\) and scale parameter \(\lambda\). Instead of ordinary least squares, we consider the least absolute deviation (LAD) loss function, \(f(\beta, \zeta_i) = {\big|y_i - x_i^\top \beta\big|}\). \cite{sgd:bootstrap} consider this simulation setup as well and discuss that the true $\Sigma$ is $A^{-1}$. 

\begin{figure}
    \centering
    \includegraphics[width = 1.9in]{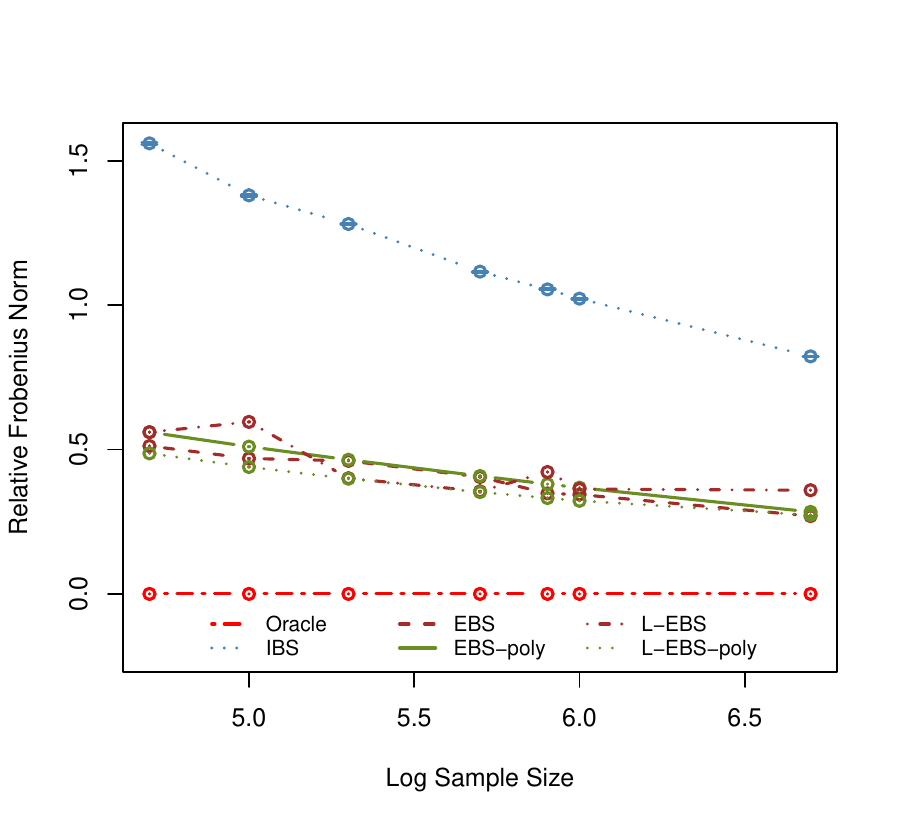}
    \includegraphics[width = 1.9in]{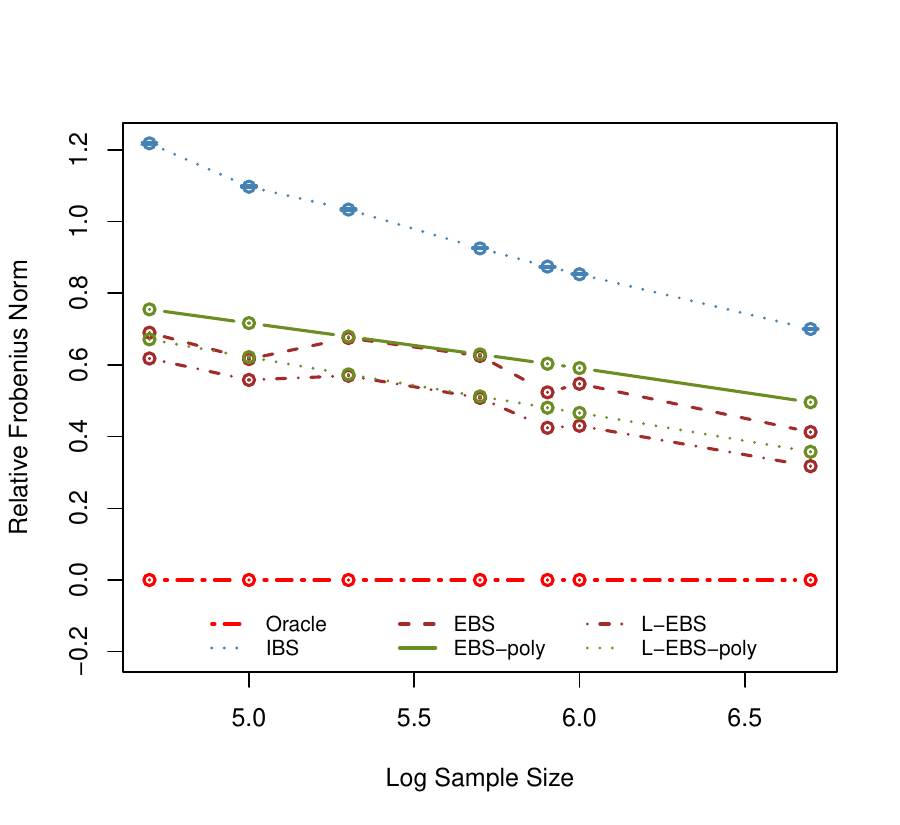}
    \includegraphics[width = 1.9in]{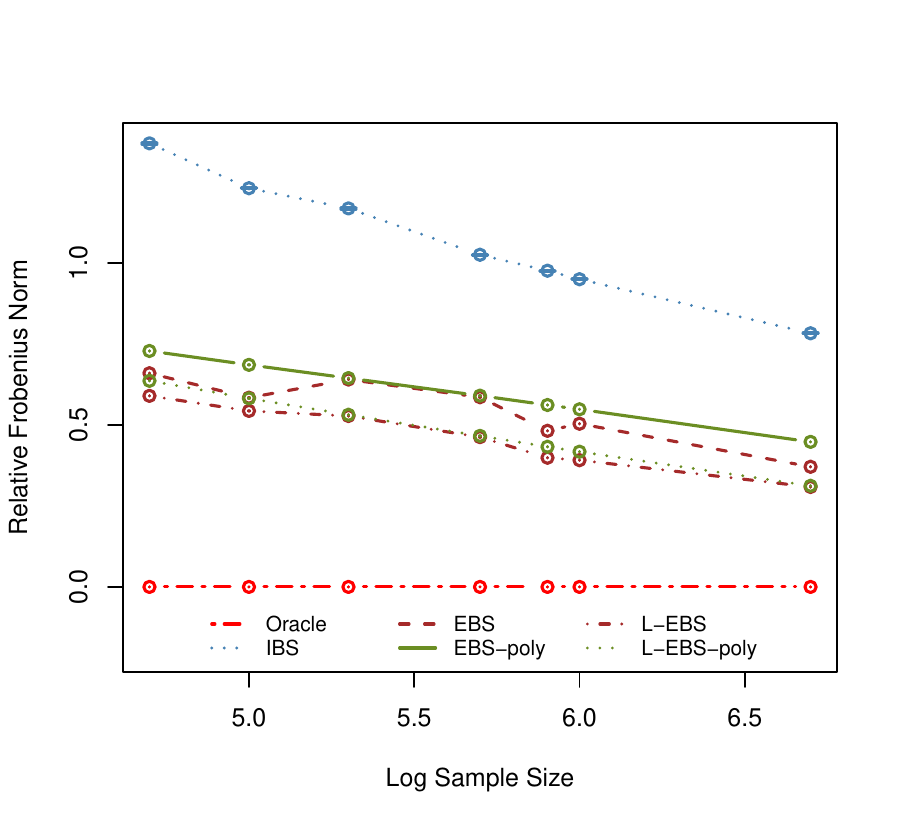}

    \includegraphics[width = 1.9in]{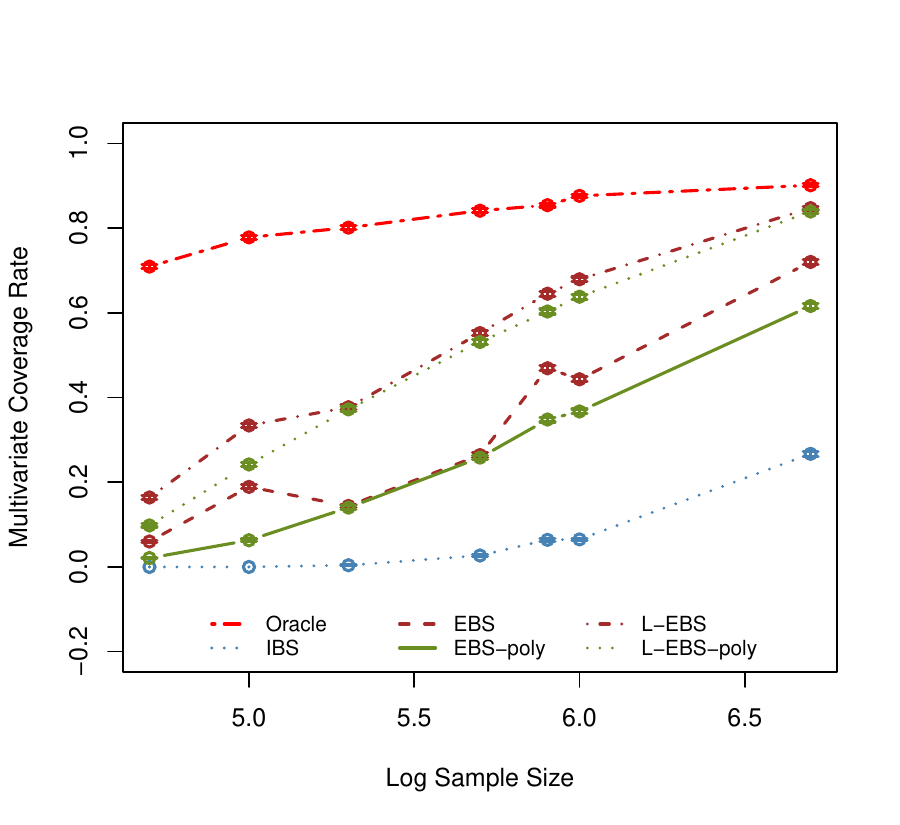}
    \includegraphics[width = 1.9in]{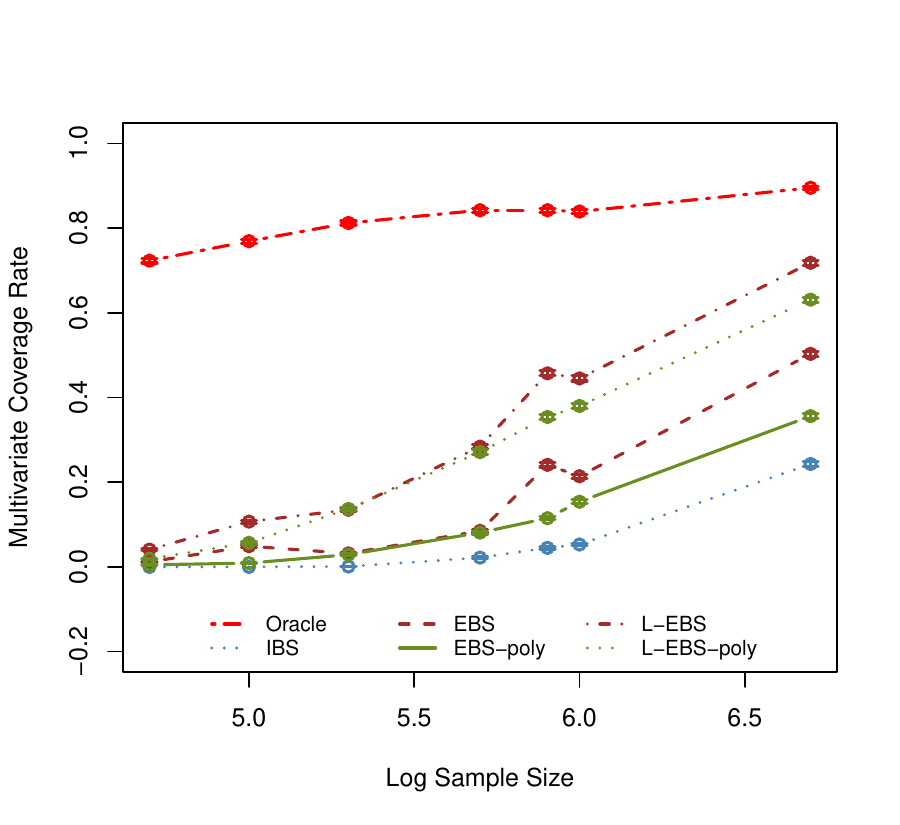}
    \includegraphics[width = 1.9in]{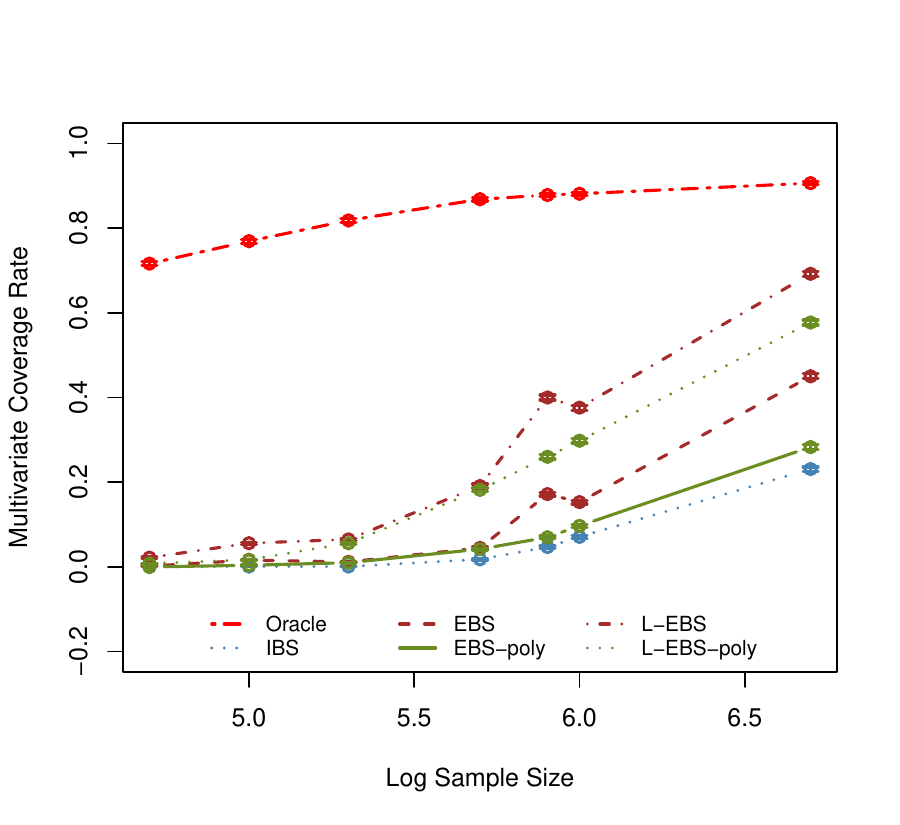}

    \includegraphics[width = 1.9in]{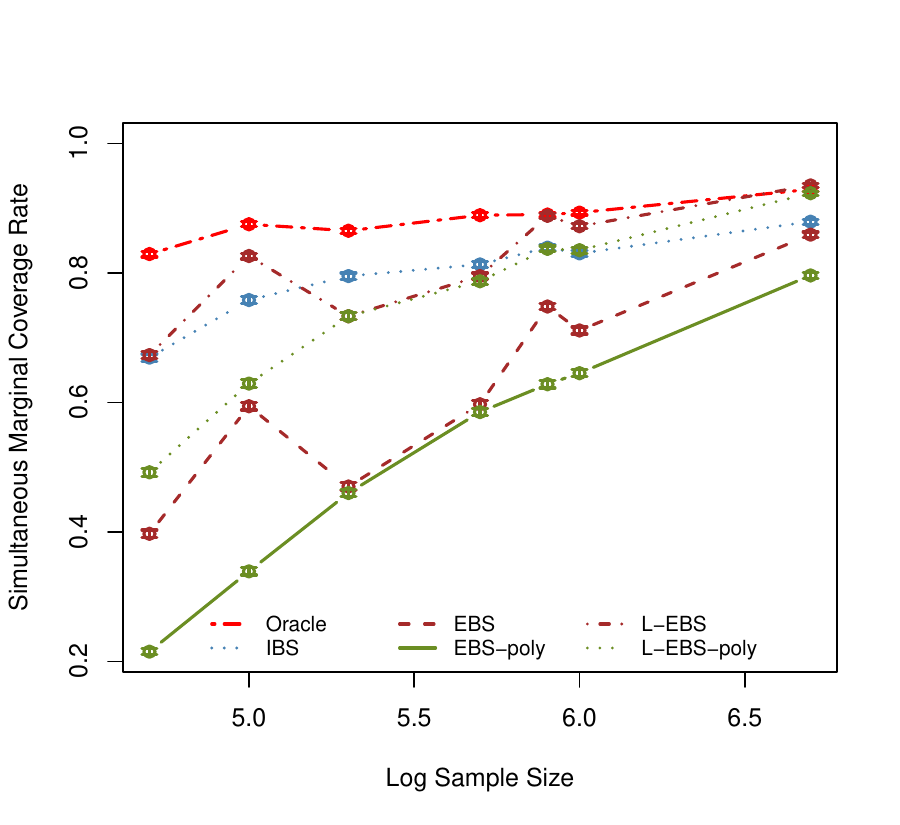}
    \includegraphics[width = 1.9in]{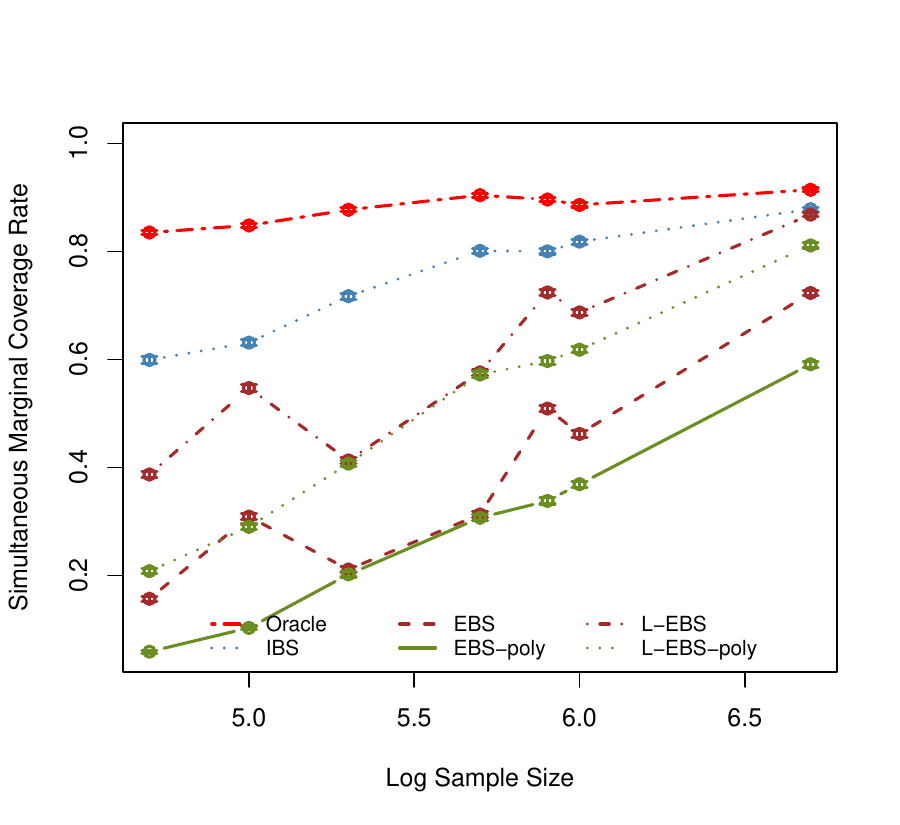}
    \includegraphics[width = 1.9in]{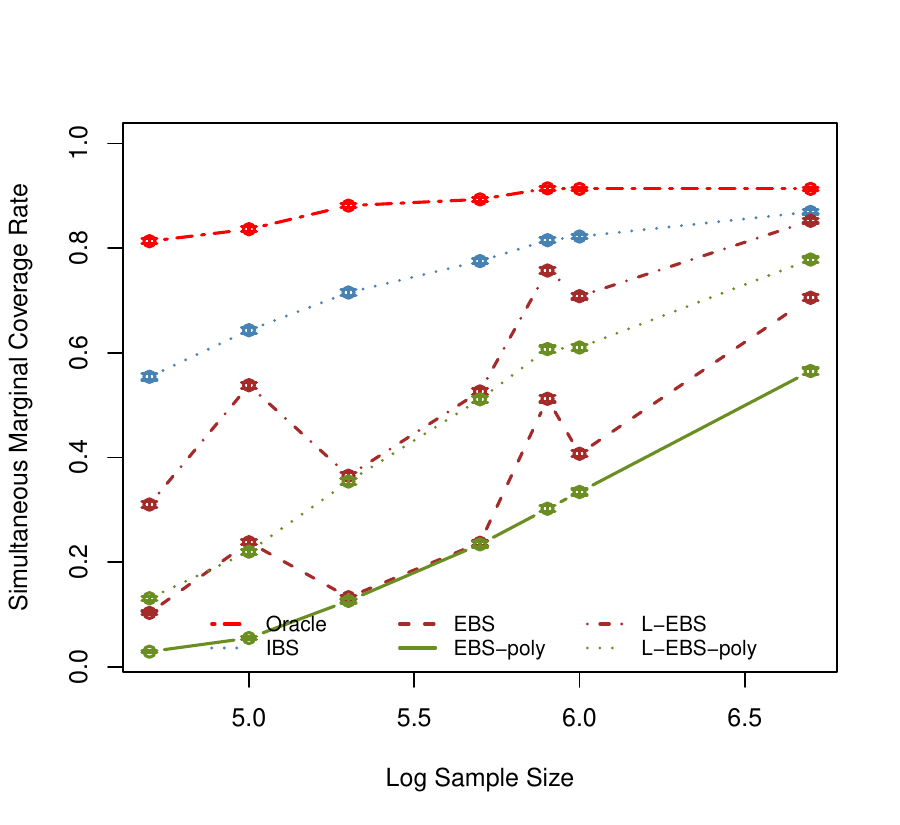}

    \includegraphics[width = 1.9in]{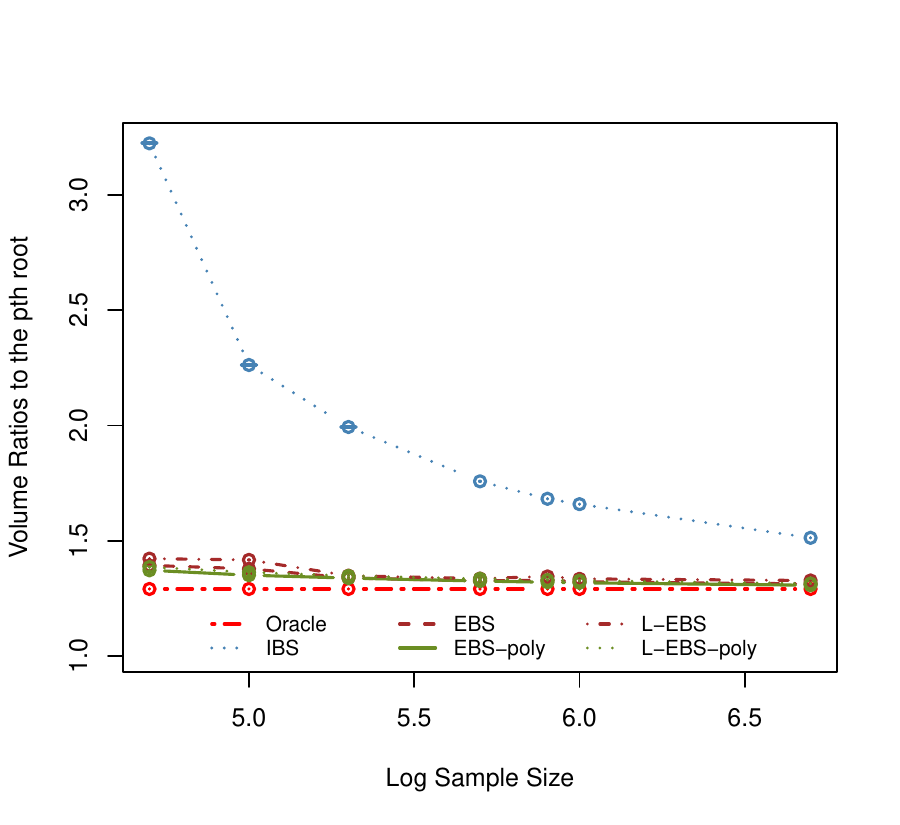}
    \includegraphics[width = 1.9in]{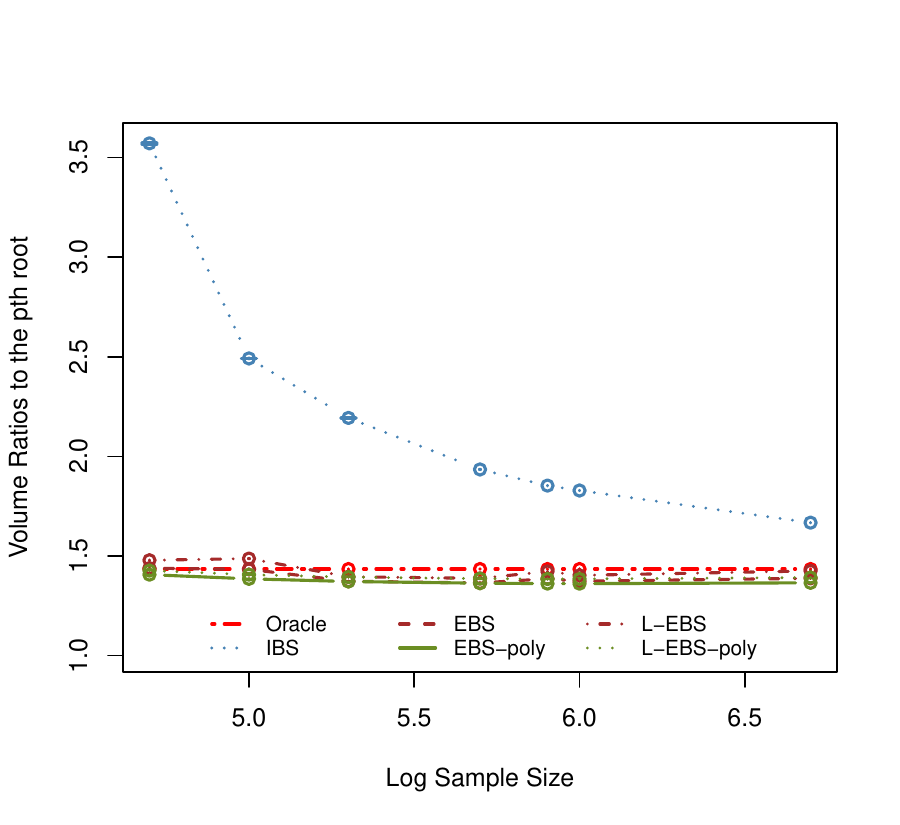}
    \includegraphics[width = 1.9in]{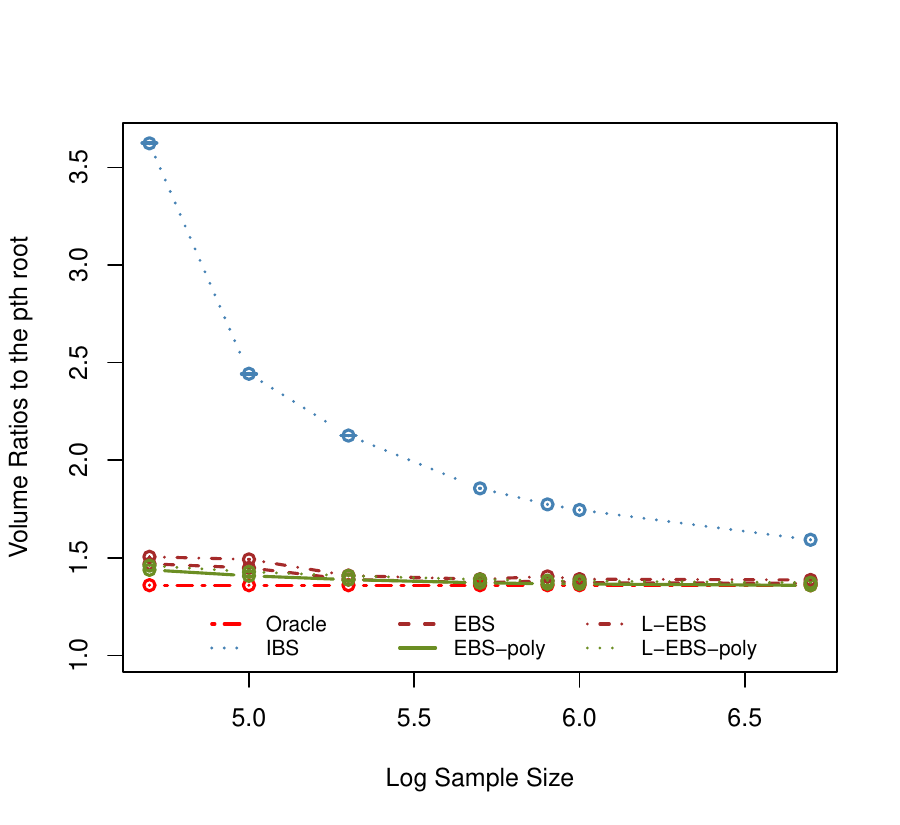}
    
\caption{LAD regression for $d = 20$: Line plots equipped with error bars from 1000 replications. Left plots for $A$ being identity, middle plots for $A$ being Toeplitz, and right plots for $A$ being equicorrelation matrix. }
\label{fig:Lad_p20}    
\end{figure}

We repeat the simulation setup of the previous section with $d = 20$ for the three different choices of $A$.  Figure~\ref{fig:Lad_p20} presents the results. The performance of the EBS estimators, particularly lugsail-EBS is significantly superior to that of the IBS estimator. Here again, although the simultaneous coverage of the hyper-rectangular regions is far improved for the IBS estimator, this is purely a consequence of over-inflated volume of the region. 	

\subsection{Improving predictions with estimators of $\Sigma$}
\label{sec:logistic}

{ Consider the binary classification problem where} for $i = 1, 2, \dots, $
\[
y_i \overset{\text{ind}}{\sim}  \text{Bernoulli} \left( p_i := \dfrac{1}{1 + e^{-x_i^{\top} \beta^*}}\right),
\]
{ with}  $x_i { \in \mathbb{R}^d}$  assumed to be iid. 
For estimation, the loss function is the negative log-likelihood as a consequence of the Bernoulli model assumption. { However,} estimates of $\beta^*$ {  that minimize this loss function} are used in predictions, without any focus on accounting for the variability is its estimation. Denote the ASGD estimator of $\beta^*$ with $\hat{\beta}_n$, and for any data point $j$, the fitted/predicted probability of success is estimated with
\[
	\hat{p}_j := \dfrac{1}{1 + e^{-x_j^{\top} \hat{\beta}_n}}\,.
\] 
A thresholding is then typically used to obtain a binary prediction of this $j$th observation, based on $\hat{p}_j$. That is, for some user-chosen threshold $q$, $\hat{y}_j = \mathbb{I}(\hat{p}_j > q)$. When employing a test dataset, misclassification rates can then be obtained for model building and comparisons. Due to \eqref{eq:pj_normal} and the delta method, as $n \to \infty$,
\[
\sqrt{n}(\hat{p}_j - p_j) \overset{d}{\to} N\left( 0, (p_j(1 - p_j))^2 x_j^{\top} \Sigma x_j  \right)\,.
\]
Since our proposed EBS estimator of $\Sigma$ is consistent, we can obtain a confidence interval for each $\hat{p}_j \pm z_{.975} \,\, \text{se}_j$, where {  $\text{se}_j$ is the standard error for the $j$th response and} is calculated using the plug-in estimators of $\Sigma$ and $p_j$. In order to account for the estimation variability in $\hat{p}_j$, we employ an alternative estimator of $y_j$: $\tilde{y}_j = \mathbb{I}(\hat{p}_j -  z_{.975} \,\, \text{se}_j > q)$. That is, our logistic classifier, classifies the observation as a success if the lower-bound on the confidence interval for $\hat{p}_j$ is larger than the cutoff, $q$; different observations will have different se$_j$.

{ We implement this strategy on four datasets (i) Santander customer transcation dataset\footnote{\url{https://www.kaggle.com/competitions/santander-customer-transaction-prediction/overview}{www.kaggle.com/competitions/santander-customer-transaction-prediction/overview}}, (ii) Covertype dataset of \cite{blackardcover}, (iii) Spambase dataset of \cite{hopkinsspam}, and (iv) the diabetes health dataset\footnote{\url{ https://archive.ics.uci.edu/dataset/891/cdc+diabetes+health+indicators}}. Implementation details for each of them is provided in Supplement E. For each dataset, since the true $\beta^*$ and $\Sigma$ are unknown, comparison of confidence regions is unreasonable here. We can obtain marginal simultaneous confidence intervals, however due to the nature of the data, classical hypothesis testing may not be of interest. Instead, we utilize the estimator of $\Sigma$ for prediction.
}

\begin{figure}[!htb]
	\centering
	\includegraphics[width = 2.5in]{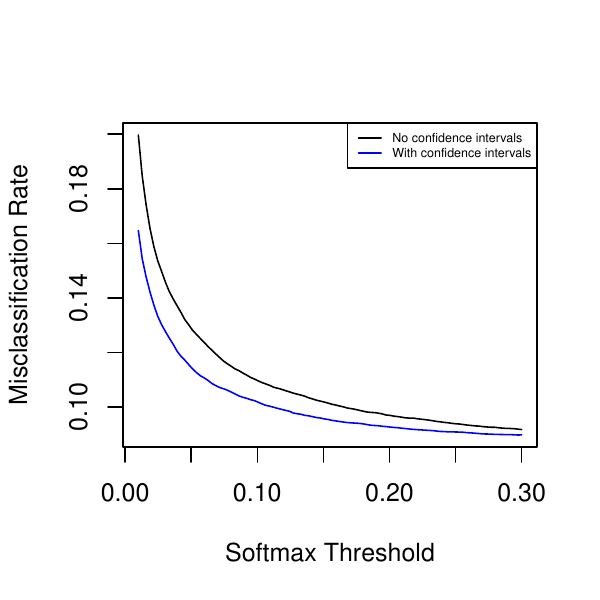}
    \includegraphics[width = 2.5in]{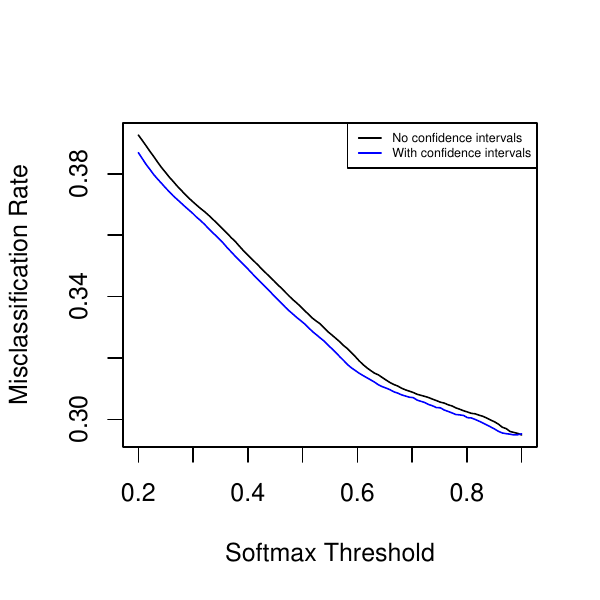}
    \includegraphics[width = 2.5in]{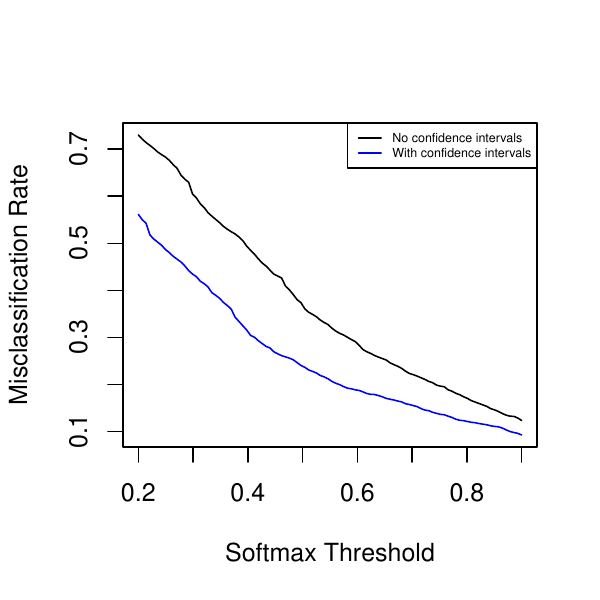}
    \includegraphics[width = 2.5in]{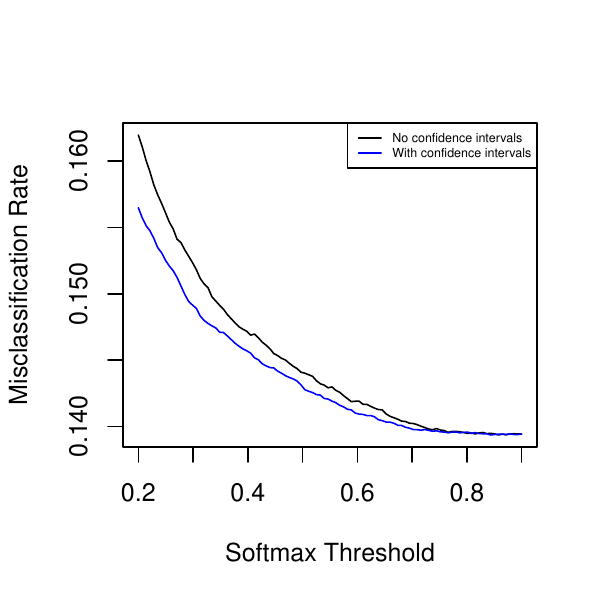}
	\caption{Misclassification rate on the testing set { for various values of the cutoff for (i) Santander (topleft) (ii) covertype (topright) (iii) spambase (bottomleft) (iv) diabetes(bottomright) datasets.}}
	\label{fig:logistic_misclass}
\end{figure}

Figure~\ref{fig:logistic_misclass} demonstrates the test data misclassification rate for various values of $q$, {  with and without confidence interval, for all datasets}. Clearly, the blue curve which employs the EBS estimator to obtain se$_j$ yields a lower misclassification rate. Due to consistency of the estimator, se$_j$ is expected to converge to 0 as $n \to \infty$, and thus, for a large enough training data, we would expect the blue and the black curves to merge into one, {  for all datasets.}

\section{Discussion}

{ We present EBS batching-strategies for batch-means estimators in order to estimate the limiting variance of SGD estimates.} Our proposed EBS batching-strategy can be extended to averaging over $k$-neighbouring batches for any fixed positive integer $k$, and all the theoretical results discussed will hold true. However, large values of $k$ will reduce the number of batches, thereby reducing the efficiency of the covariance estimator. { Another alternative is to adopt an overlapping batch-means estimator with an EBS strategy, adapting  the IBS estimator of \cite{chen2021jasa}. Overlapping batch-means estimators have found reasonable success in stochastic simulation \citep{meketon1984overlapping} as they allow higher number of batches, ableit correlated. However, the computational complexity of these estimators is $\mathcal{O}(d^2n)$ since the number of batches are on the order of the number of samples. Nonetheless, similar theoretical results should be possible for overlapping batch-means estimators with an EBS strategy.} \cite{leung:chan:2022} discuss variants of the IBS estimator and find that their performances are quite similar. A study similar to \cite{leung:chan:2022} for the EBS estimator would make a useful follow-up of our work. 	
Building a statistical inference framework for SGD is an active area of research in recent times. This includes the recent works of \cite{sgd:bootstrap,xie2023scalable} who use bootstrap techniques to estimate the limiting covariance structure. \cite{sgd:bootstrap} heuristically argued to use the (perturbed) bootstrap ASGD outputs to estimate the covariance matrix of $\hat{\theta}_n$. The theoretical properties of the estimator are not known and computational demands of the estimator is considerable. \cite{dong2021} present a method of consistent inference for SGD without using a consistent estimator of $\Sigma$. It remains unclear if methods of marginal inference and delta method arguments can be used in their framework. \cite{li2023online,liu2023statistical} estimate the limiting covariance under situations when the iid assumption on the data is violated. Our marginal-friendly confidence interval construction, and utilization of $\Sigma$ in improving predictions are directly applicable to this literature. 

There are numerous other variants of the SGD \citep[see, e.g., ][]{minibatch:sgd,implicit:sgd,momentum:sgd,accelerted:sgd}, and the fundamental framework remains essentially the same for other variants of SGD, as long as the results of \cite{pj1992} applies to them.   For example, \cite{implicit:sgd} obtained asymptotic normality of averaged implicit SGD, and the framework we present here can be seamlessly transferred to that setup. 

Finally, we employ the estimator of $\Sigma$ in two tasks: (i) the construction of marginal-friendly simultaneous confidence intervals that favor interpretability over ellipsoidal regions, and (ii) construct confidence intervals around predictions for new observations. The { classification} example we present in Section~\ref{sec:logistic} demonstrates this feature. A similar argument can yield prediction intervals for regression as well, one which accounts for the multivariate estimation error in the SGD estimates. 

\section*{Acknowledgements} 
\label{sub:acknowledgements}
The authors sincerely appreciate the time and effort of two anonymous Reviewers in reviewing our paper and providing insightful feedback, which has significantly improved the presentation of our work. 
The authors are thankful to Prof. Jing Dong for useful conversations. Dootika Vats is supported by SERB (SPG/2021/001322) and Google Asia Pacific Pte Ltd.

\begin{center}
	{\large SUPPLEMENTARY MATERIAL}
\end{center}
\begin{enumerate}
	\item {\bf Supplement to ``On the Utility of Equal Batch Sizes for Inference in  Stochastic Gradient Descent'':} This contains some additional lemmas, technical proofs of the results, and some additional details on the numerical studies presented in the main text (attached herewith).
	
	\item {\bf Reproducible codes:} All relevant codes are provided { in the following Github repository: \url{https://github.com/Abhinek-Shukla/SGD-EBS}.}
\end{enumerate}

\bibliography{sgd}

\newpage



\section*{Supplementary material}

4. For any fixed $\gamma > 0$, $\sum_{k=1}^{T} k^{\gamma} \leq \int_{1}^{T}(x+1)^{\gamma}\, dx \lesssim T^{1+\gamma}$.

*Proof.* Proofs of 1) and 2) are obvious. For the proof of 3), observe that

$$T^{-1} \sum_{k=1}^{T} k^{-\gamma} \leq T^{-1}\left[1 + \int_{1}^{T+1} x^{-\gamma} dx\right] \lesssim T^{-1} \int_{1}^{T+1} x^{-\gamma} dx \lesssim T^{-\gamma}.$$

Next, for the proof of 4), we have

$$\sum_{k=1}^{T} k^{\gamma} \leq \int_{1}^{T}(x+1)^{\gamma}\, dx \leq \frac{1}{1+\gamma}(T+1)^{\gamma+1} \lesssim T^{1+\gamma}.$$

□

The following three results from Chen et al. (2020) will be used in our proofs.

**Result 1** (Chen et al. (2020), Lemma D.2). *For each positive integer $j$, let $Y_j^k$ be a sequence of matrices with $Y_j^j = I$ and for $k \geq j$, $Y_j^k = (I - \eta_k A)Y_j^{k-1} = \prod_{i=j}^{k}(I - \eta_i A)$, where $A$ is a PSD matrix with eigen values bounded below by $\lambda_A > 0$. Then, under the assumption (A3), we have*

1) *For $i \in \{j, \ldots, k\}$, the following holds*

$$\|Y_j^k\| \leq \exp\left(-\lambda_A \sum_{i=j+1}^{k} \eta_i\right) \leq \exp\left(-\lambda_A(k-j)\eta_k\right).$$

2) *Let $S_j^k = \sum_{i=j+1}^{k} Y_j^i$ and $Z_j^k = \eta_j S_j^k - A^{-1}$, then*

$$\|S_j^k\| \lesssim k^{\alpha},\ \|Z_j^k\| \lesssim k^{\alpha} j^{-1} + \exp\left(-\lambda_A \sum_{i=j}^{k} \eta_i\right).$$

3) *When $1 \leq j < k \leq \tau_{a_n}$, for sufficiently large $N$, there exists a constant $c$ such that*

$$\|S_j^k\| \lesssim j^{\alpha},\ \|Z_j^k\| \lesssim j^{\alpha-1} + \exp\left(-c\lambda_A(k-j)\eta_j\right),\ \|\eta_j S_j^k\| \lesssim 1,$$

*as long as $a_n \leq N^q$ for certain positive integer $q$.*

25

**Result 2** (Chen et al. (2020), Lemma D.3). *Let $S$ and $\Sigma$ are two $d \times d$ matrices, and $S$ is positive definite, then $\operatorname{tr}(S\Sigma) \leqslant \operatorname{tr}(S)\|\Sigma\|$.*

**Result 3** (Chen et al. (2020), Lemma 3.2). *Under the assumptions (A1) and (A2), there exists $n_0$ such that the iterate error $D_n$ satisfy the following:*

1) *For $n > m \geqslant n_0$,*

$$\mathbb{E}_m \|D_n\| \lesssim \exp\left(\tfrac{-1}{4} M \sum_{i=m}^{n} \eta_i\right) \|D_m\| + \sqrt{C_d} m^{-\alpha/2},$$

$$\mathbb{E}_m \|D_n\|^2 \lesssim \exp\left(\tfrac{-1}{4} M \sum_{i=m}^{n} \eta_i\right) \|D_m\|^2 + C_d m^{-\alpha}, \text{ and}$$

$$\mathbb{E}_m \|D_n\|^4 \lesssim \exp\left(\tfrac{-1}{4} M \sum_{i=m}^{n} \eta_i\right) \|D_m\|^4 + C_d^2 m^{-2\alpha}.$$

2) $\quad \mathbb{E}\|D_n\| \lesssim n^{-\alpha/2}(\sqrt{C_d} + \|D_{n_0}\|),$

$\quad \mathbb{E}\|D_n\|^2 \lesssim n^{-\alpha}(C_d + \|D_{n_0}\|^2), \text{ and}$

$\quad \mathbb{E}\|D_n\|^4 \lesssim n^{-2\alpha}(C_d^2 + \|D_{n_0}\|^4).$

# B  Auxiliary sequence

An approximation of the SGD iterate is $D_i \simeq (I_d - \eta_i A) D_{i-1} + \eta_i \xi_i$. We replace $\simeq$ by $=$ to obtain an approximated iterate sequence,

$$U_i = (I_d - \eta_i A) U_{i-1} + \eta_i \xi_i, \ U_0 = D_0. \tag{17}$$

The sequence $\{U_i\}$ is known as the auxiliary sequence. The following result for the sequence $\{U_i\}$ from Chen et al. (2020) is helpful.

**Result 4** (Chen et al. (2020), Lemma B.3). *For the sequence $U_i$, under the assumptions (A1) and (A2), we have*

$$\mathbb{E}\|U_n\|^2 \lesssim n^{-\alpha}(C_d + \|U_0\|^2 + \|D_0\|^2).$$

**Lemma 2.** *Under the assumptions (A1), (A2) and (A3), we have*

$$\hat{S} := \frac{1}{a_n} \sum_{k=1}^{a_m} \frac{1}{b_n} \left( \sum_{i=\tau_{k-1}+1}^{\tau_k} \xi_i \right) \left( \sum_{i=\tau_{k-1}+1}^{\tau_k} \xi_i \right)^\top$$

*is a consistent estimator of $S$, precisely,*

$$\mathbb{E}\|\hat{S} - S\|^2 \lesssim C_d^4 n^{-\alpha+(\beta-1)/2} + C_d^6 n^{-2\alpha} + C_d^2 a_n^{-1}.$$

*Proof.* Let $\tilde{\xi}_i = -\Delta f(\theta^*, \zeta_i)$, $\hat{\xi}_i = \xi_i - \tilde{\xi}_i$ and

$$\tilde{S} = \frac{1}{a_n} \sum_{k=1}^{a_n} \frac{1}{b_n} \left( \sum_{i=\tau_{k-1}+1}^{\tau_k} \tilde{\xi}_i \right) \left( \sum_{i=\tau_{k-1}+1}^{\tau_k} \tilde{\xi}_i \right)^\top.$$

Then, $\xi_i = \hat{\xi}_i + \tilde{\xi}_i$ where $\{\tilde{\xi}_i\}$ is a sequence of iid RVs and $\{\hat{\xi}_i\}$ is a martingale difference sequence. Note that

$$\mathbb{E}\|\hat{S} - S\|^2 \lesssim \mathbb{E}\|\tilde{S} - S\|^2 + \mathbb{E}\|\hat{S} - \tilde{S}\|^2, \tag{18}$$

and

$$\mathbb{E}\|\tilde{S} - S\|^2 \leqslant \mathbb{E}\,\mathrm{tr}(\tilde{S} - S)^2 = \mathrm{tr}\,\mathbb{E}(\tilde{S} - S)^2. \tag{19}$$

Further

$$\mathbb{E}(\tilde{S}) = \frac{1}{a_n} \sum_{k=1}^{a_n} \frac{1}{b_n} \sum_{i=\tau_{k-1}+1}^{\tau_k} \mathbb{E}(\tilde{\xi}_i \tilde{\xi}_i^\top) = S,$$

therefore $\mathbb{E}(\tilde{S} - S)^2 = \mathbb{E}(\tilde{S}^2) - S^2$. Next

$$\mathbb{E}\tilde{S}^2 = a_n^{-2} \sum_{j=1}^{a_n} \sum_{k=1}^{a_n} b_n^{-2} \sum_{i_1, i_2 = \tau_{j-1}+1}^{\tau_j} \sum_{i_3, i_4 = \tau_{k-1}+1}^{\tau_k} \mathbb{E}(\tilde{\xi}_{i_1} \tilde{\xi}_{i_2}^\top \tilde{\xi}_{i_3} \tilde{\xi}_{i_4}^\top).$$

27

Since $\{\widetilde{\xi}_i\}$ is an iid sequence of RVs, on the RHS of the above expression, the terms are zero unless $i_1 = i_2$, $i_3 = i_4$ or $i_1 = i_3$, $i_2 = i_4$ or $i_1 = i_4$, $i_2 = i_3$. Also, the latter two only happen when all the indices are in the same batch, thus

$$\mathbb{E}\widetilde{S}^2 = a_n^{-2} \sum_{j=1}^{a_n} \sum_{k=1}^{a_n} b_n^{-2} \sum_{i_1=\tau_{j-1}+1}^{\tau_j} \sum_{i_3=\tau_{k-1}+1}^{\tau_k} \mathbb{E}(\widetilde{\xi}_{i_1}\widetilde{\xi}_{i_1}^\top \widetilde{\xi}_{i_3}\widetilde{\xi}_{i_3}^\top)$$
$$+ a_n^{-2} \sum_{j=1}^{a_n} b_n^{-2} \sum_{i_1,i_3=\tau_{j-1}+1, i_1\neq i_3}^{\tau_j} [\mathbb{E}(\widetilde{\xi}_{i_1}\widetilde{\xi}_{i_3}^\top \widetilde{\xi}_{i_1}\widetilde{\xi}_{i_3}^\top) + \mathbb{E}(\widetilde{\xi}_{i_1}\widetilde{\xi}_{i_3}^\top \widetilde{\xi}_{i_3}\widetilde{\xi}_{i_1}^\top)].$$

Using $\mathbb{E}(\widetilde{\xi}_{i_1}\widetilde{\xi}_{i_1}^\top \widetilde{\xi}_{i_3}\widetilde{\xi}_{i_3}^\top) = S^2$ if $i_1 \neq i_3$, we get

$$\operatorname{tr}(\mathbb{E}\widetilde{S}^2 - S^2) = a_n^{-2} \sum_{j=1}^{a_n} b_n^{-2} \sum_{i_1,i_3=\tau_{j-1}+1, i_1\neq i_3}^{\tau_j} [\mathbb{E}(\widetilde{\xi}_{i_1}\widetilde{\xi}_{i_3}^\top \widetilde{\xi}_{i_1}\widetilde{\xi}_{i_3}^\top) + \mathbb{E}(\widetilde{\xi}_{i_1}\widetilde{\xi}_{i_3}^\top \widetilde{\xi}_{i_3}\widetilde{\xi}_{i_1}^\top)].$$

Next using $\mathbb{E}(\widetilde{\xi}_{i_1}\widetilde{\xi}_{i_1}^\top \widetilde{\xi}_{i_3}\widetilde{\xi}_{i_3}^\top) \lesssim C_d^2 + C_d^6 i_1^{-2\alpha}$ if $i_1 = i_3$, we obtain

$$\operatorname{tr}(\mathbb{E}\widetilde{S}^2 - S^2) = a_n^{-2} \sum_{j=1}^{a_n} b_n^{-2} \left( \sum_{i_1=\tau_{j-1}+1}^{\tau_j} \mathbb{E}\|\widetilde{\xi}_{i_1}\|^4 + \sum_{i_1,i_3=\tau_{j-1}+1, i_1\neq i_3}^{\tau_j} (\mathbb{E}(\widetilde{\xi}_{i_1}^\top \widetilde{\xi}_{i_3})^2 + \mathbb{E}(\|\widetilde{\xi}_{i_1}\|^2 \|\widetilde{\xi}_{i_3}\|^2)) \right)$$
$$\lesssim a_n^{-2} \sum_{j=1}^{a_n} \left( b_n^{-1}(C_d^2 + C_d^6 \tau_j^{-2\alpha}) + (\operatorname{tr}(S))^2 \right).$$

Consequently using Lemma 1

$$\operatorname{tr}(\mathbb{E}\widetilde{S}^2 - S^2) \lesssim C_d^2 n^{-1} + C_d^6 a_n^{-(1+2\alpha)} + C_d^2 a_n^{-1}. \tag{20}$$

Next, we denote

$$\check{S} = \frac{1}{a_n} \sum_{k=1}^{a_n} \frac{1}{b_n} \left( \sum_{i=\tau_{k-1}+1}^{\tau_k} \hat{\xi}_i \right) \left( \sum_{i=\tau_{k-1}+1}^{\tau_k} \hat{\xi}_i \right)^\top.$$

We notice that

$$\mathbb{E}\|\hat{S}-\widetilde{S}\|^2 = \mathbb{E}\left\|a_n^{-1}\sum_{k=1}^{a_n}b_n^{-1}\left(\left(\sum_{i=\tau_{k-1}+1}^{\tau_k}\widetilde{\xi}_i\right)\left(\sum_{i=\tau_{k-1}+1}^{\tau_k}\hat{\xi}_i\right)^\top + \left(\sum_{i=\tau_{k-1}+1}^{\tau_k}\hat{\xi}_i\right)\left(\sum_{i=\tau_{k-1}+1}^{\tau_k}\widetilde{\xi}_i\right)^\top\right) + 2\breve{S}\right\|^2$$

$$\lesssim \mathbb{E}\left\|a_n^{-1}\sum_{k=1}^{a_m}b_n^{-1}\left(\sum_{i=\tau_{k-1}+1}^{\tau_k}\widetilde{\xi}_i\right)\left(\sum_{i=\tau_{k-1}+1}^{\tau_k}\hat{\xi}_i\right)^\top\right\|^2 + \mathbb{E}\|\breve{S}\|^2$$

$$\lesssim \sqrt{\mathbb{E}(\mathrm{tr}(\breve{S}^2))}\sqrt{\mathbb{E}(\mathrm{tr}(\widetilde{S}^2))} + \mathbb{E}(\mathrm{tr}(\breve{S}^2)). \tag{21}$$

Now we need to find a bound for $\mathbb{E}(\mathrm{tr}(\breve{S}^2))$. Let $\bar{\xi}_k = b_n^{-1}\sum_{i=\tau_{k-1}+1}^{\tau_k}\hat{\xi}_i$. Then using Cauchy-Schwartz inequality, we have

$$\mathbb{E}(\mathrm{tr}(\breve{S}^2)) = a_n^{-2}\sum_{k=1}^{a_n}\sum_{l=1}^{a_n}b_n^2\,\mathbb{E}(\mathrm{tr}(\bar{\xi}_k\bar{\xi}_k^\top\bar{\xi}_l\bar{\xi}_l^\top)) \lesssim a_n^{-2}\sum_{k=1}^{a_n}\sum_{l=1}^{a_n}b_n^2\sqrt{\mathbb{E}\|\bar{\xi}_k\|^4}\sqrt{\mathbb{E}\|\bar{\xi}_l\|^4}$$

$$\lesssim a_n^{-2}\sum_{k=1}^{a_n}\sum_{l=1}^{a_n}b_n^{-2}\sum_{i=\tau_{k-1}+1}^{\tau_k}\sqrt{\mathbb{E}\|\hat{\xi}_i\|^4}\sum_{j=\tau_{l-1}+1}^{\tau_l}\sqrt{\mathbb{E}\|\hat{\xi}_j\|^4},\text{ using Theorem 2.1 in Rio (2009).}$$

Using Result 3, we have $\mathbb{E}\|\hat{\xi}_j\|_2^4 \lesssim C_d^6 i^{-2\alpha}$. Consequently using Lemma 1 we get

$$\mathbb{E}(\mathrm{tr}(\breve{S}^2)) \lesssim C_d^6(n^{-1}\sum_{i=1}^n i^{-\alpha})(n^{-1}\sum_{j=1}^n j^{-\alpha}) \lesssim C_d^6 n^{-2\alpha}.$$

Further substituting in (21) we have

$$\mathbb{E}\|\hat{S}-\widetilde{S}\|^2 \lesssim C_d^4 n^{-\alpha+(\beta-1)/2} + C_d^6 n^{-2\alpha}. \tag{22}$$

Combining (18), (19), (20), (21) and (22), completes the proof. $\square$

Define overall mean and batch means of $U_i$ as follows:

$$\bar{U}_n = \frac{1}{\tau_{a_n}}\sum_{i=1}^{\tau_{a_n}}U_i \text{ and } \bar{U}_k = \frac{1}{b_n}\sum_{i=\tau_{k-1}+1}^{\tau_k}U_i \text{ for } k=1,2,\ldots,a_n.$$

29

**Lemma 3.** *Under the assumptions (A1), (A2) and (A3), we have*

$$\mathbb{E}\left\|a_n^{-1}\sum_{k=1}^{a_n} b_n \bar{U}_k \bar{U}_k^\top - A^{-1}SA^{-1}\right\| \lesssim C_d^2 n^{-\alpha/2+(\beta-1)/4} + C_d^3 n^{-\alpha} + C_d a_n^{-1/2} + C_d b_n^{\alpha-1} + C_d b_n^{-1/2} n^{\alpha/2}.$$

*Proof.* The recursion of $U_j$ can be written as

$$U_i = Y_{i-1}^i U_{i-1} + Y_i^i \eta_i \xi_i = Y_{i-2}^i U_{i-2} + Y_{i-1}^i \eta_{i-1} \xi_{i-1} + Y_i^i \eta_i \xi_i$$

$$= Y_j^i U_j + \sum_{l=j+1}^{i} Y_l^i \eta_l \xi_l.$$

Therefore, the $k^{th}$ batch mean $\bar{U}_k$ can be written as

$$\bar{U}_k = b_n^{-1}\left(\sum_{i=\tau_{k-1}+1}^{\tau_k} Y_{\tau_{k-1}}^i U_{\tau_{k-1}} + \sum_{i=\tau_{k-1}+1}^{\tau_k}\sum_{l=\tau_{k-1}+1}^{i} Y_l^i \eta_l \xi_l\right)$$

$$= b_n^{-1}\left[\sum_{i=\tau_{k-1}+1}^{\tau_k} Y_{\tau_{k-1}}^i U_{\tau_{k-1}} + \sum_{l=\tau_{k-1}+1}^{\tau_k}\left(\sum_{i=l}^{\tau_k} Y_l^i\right)\eta_l \xi_l\right]$$

$$= b_n^{-1}\left[\sum_{i=\tau_{k-1}+1}^{\tau_k} Y_{\tau_{k-1}}^i U_{\tau_{k-1}} + \sum_{l=\tau_{k-1}+1}^{\tau_k}\left(S_l^{\tau_k} + I\right)\eta_l \xi_l\right]$$

$$= b_n^{-1} S_{\tau_{k-1}}^{\tau_k} U_{\tau_{k-1}} + b_n^{-1} A^{-1}\sum_{l=\tau_{k-1}+1}^{\tau_k}\xi_l + b_n^{-1}\sum_{l=\tau_{k-1}+1}^{\tau_k}\xi_l Z_l^{\tau_k} + b_n^{-1}\sum_{l=\tau_{k-1}+1}^{\tau_k}\eta_l\xi_l.$$

Let us denote

$$A_k := -A^{-1}\sum_{l=\tau_{k-1}+1}^{\tau_k}\xi_l \quad \text{and} \quad B_k := S_{\tau_{k-1}}^{\tau_k} U_{\tau_{k-1}} + \sum_{l=\tau_{k-1}+1}^{\tau_k}\xi_l Z_l^{\tau_k} + \sum_{l=\tau_{k-1}+1}^{\tau_k}\eta_l\xi_l.$$

Then,

$$a_n^{-1}\sum_{k=1}^{a_n} b_n \bar{U}_k \bar{U}_k^\top = a_n^{-1}\sum_{k=1}^{a_n} b_n^{-1} A_k A_k^\top + a_n^{-1}\sum_{k=1}^{a_n} b_n^{-1}[A_k B_k^\top + B_k A_k^\top + B_k B_k^\top]. \tag{23}$$

We have from Lemma 2,

$$\mathbb{E}\left\|a_n^{-1}\sum_{k=1}^{a_n} b_n^{-1} A_k A_k^\top - A^{-1}SA^{-1}\right\|^2 = \mathbb{E}\left\|A^{-1}(\hat{S}-S)A^{-1}\right\|^2$$

$$\lesssim C_d^4 n^{-\alpha+(\beta-1)/2} + C_d^6 n^{-2\alpha} + C_d^2 a_n^{-1}. \qquad (24)$$

Using Cauchy Schwartz inequality, we have

$$\mathbb{E}\left\|a_n^{-1}\sum_{k=1}^{a_n} b_n^{-1} A_k A_k^\top - A^{-1}SA^{-1}\right\| \lesssim C_d^2 n^{-\alpha/2+(\beta-1)/4} + C_d^3 n^{-\alpha} + C_d a_n^{-1/2}.$$

Using $\|B_k\|^2 = \|B_k B_k^\top\| \leq \operatorname{tr}(B_k B_k^\top)$ and $\xi_i$ is a martingale, we have

$$\mathbb{E} B_k B_k^\top$$
$$= \mathbb{E}\left[S_{\tau_{k-1}}^{\tau_k} U_{\tau_{k-1}} + \sum_{l=\tau_{k-1}+1}^{\tau_k} \xi_l Z_l^{\tau_k} + \sum_{l=\tau_{k-1}+1}^{\tau_k} \eta_l \xi_l\right]\left[S_{\tau_{k-1}}^{\tau_k} U_{\tau_{k-1}} + \sum_{l=\tau_{k-1}+1}^{\tau_k} \xi_l Z_l^{\tau_k} + \sum_{l=\tau_{k-1}+1}^{\tau_k} \eta_l \xi_l\right]^\top$$
$$= S_{\tau_{k-1}}^{\tau_k} \mathbb{E}\left[U_{\tau_{k-1}} U_{\tau_{k-1}}^\top\right](S_{\tau_{k-1}}^{\tau_k})^\top + \sum_{l=\tau_{k-1}+1}^{\tau_k} (Z_l^{\tau_k} + \eta_l I)\mathbb{E}(\xi_l \xi_l^\top)(Z_l^{\tau_k} + \eta_l I)^\top.$$
$$\Rightarrow \operatorname{tr}(\mathbb{E}B_k B_k^\top) \leq \|S_{\tau_{k-1}}^{\tau_k}\|^2 \mathbb{E}\|U_{\tau_{k-1}}\|^2 + \sum_{l=\tau_{k-1}+1}^{\tau_k} \|Z_l^{\tau_k} + \eta_l I\|^2 \mathbb{E}\|\xi_l\|^2.$$

Now, using Result 1, we have

$$\|Z_l^{\tau_k} + \eta_l I\| \leq \|Z_l^{\tau_k}\| + \eta_l l^{-\alpha} \lesssim l^{\alpha-1} + \exp\left(-c\lambda_A(\tau_k - l)\eta_l\right) + l^{-\alpha}.$$

Using $l^{-\alpha} < l^{\alpha-1}$ for $\alpha \in (0.5, 1)$ and Cauchy-Schwartz inequality, we get

$$\|Z_l^{\tau_k} + \eta_l I\|^2 \lesssim l^{2\alpha-2} + \exp\left(-2c\lambda_A(\tau_k - l)\eta_l\right).$$

Next, using the bound of $\|U_n\|^2$ in Result 4, and $\mathbb{E}\|\xi_l\|^2 \lesssim C_d$, we obtain

$$\frac{1}{C_d}\operatorname{tr}\left(\mathbb{E}B_k B_k^\top\right) \lesssim (\tau_{k-1}+1)^\alpha + \sum_{l=\tau_{k-1}+1}^{\tau_k} \|Z_l^{\tau_k} + \eta_l I\|^2$$

31

$$\lesssim (\tau_{k-1}+1)^\alpha + b_n(\tau_{k-1}+1)^{2\alpha-2} + \sum_{l=\tau_{k-1}+1}^{\tau_k} \exp\left(-2c\lambda_A(\tau_k-l)\eta_{\tau_{k-1}+1}\right)$$

$$\lesssim (\tau_{k-1}+1)^\alpha + b_n(\tau_{k-1}+1)^{2\alpha-2} + \sum_{l=0}^{\infty} \exp\left(-2c\lambda_A l\eta_{\tau_{k-1}+1}\right)$$

$$\lesssim (\tau_{k-1}+1)^\alpha + b_n(\tau_{k-1}+1)^{2\alpha-2} + \left(2c\lambda_A \eta_{\tau_{k-1}+1}\right)^{-1}.$$

Using

$$b_n(\tau_{k-1}+1)^{2\alpha-2} + \left(2c\lambda_A \eta_{\tau_{k-1}+1}\right)^{-1} = b_n(k-1)^{2\alpha-2}b_n^{2\alpha-2} + (k-1)^{-\alpha}b_n^{-\alpha}\frac{1}{2c\lambda_A}$$

$$\lesssim (k-1)^{2\alpha-2}b_n^{2\alpha-1} + (k-1)^{-\alpha}b_n^{-\alpha} \lesssim b_n^{2\alpha-1},$$

and Lemma 1 (claim 4), we have

$$a_n^{-1}\sum_{k=1}^{a_n} b_n^{-1}\mathbb{E}\|B_k\|^2 \lesssim C_d a_n^{-1}\sum_{k=1}^{a_n} b_n^{-1}(b_n^{2\alpha-1} + (\tau_{k-1}+1)^\alpha)$$

$$\leqslant C_d b_n^{2\alpha-2} + C_d a_n^{-1}\sum_{k=1}^{a_n} b_n^{-1}(\tau_{k-1}+1)^\alpha$$

$$\leqslant C_d b_n^{2\alpha-2} + C_d b_n^{-1} n^\alpha. \tag{25}$$

On the other hand, using martingale property of $\xi_i$, we have

$$b_n^{-1}\mathbb{E}\|A_k\|^2 = b_n^{-1}\operatorname{tr}\left(\mathbb{E}A_k A_k^\top\right) = b_n^{-1}\operatorname{tr}\left(A^{-1}\mathbb{E}\left(\sum_{i=\tau_{k-1}+1}^{\tau_k}\xi_i\right)\left(\sum_{i=\tau_{k-1}+1}^{\tau_k}\xi_i\right)^\top A^{-1}\right)$$

$$\leqslant b_n^{-1}\|A^{-1}\|^2\left[b_n\operatorname{tr}(S) + 2\sum_{i=\tau_{k-1}+1}^{\tau_k}\mathbb{E}(\sigma_1\|D_i\| + \sigma_2\|D_i\|^2)\right] \lesssim C_d. \tag{26}$$

So, using (25), 26 and Cauchy Schwartz inequality, we get

$$b_n^{-1}\mathbb{E}\|A_k\|\|B_k\| \lesssim \sqrt{b_n^{-1}\mathbb{E}\|A_k\|^2}\sqrt{b_n^{-1}\mathbb{E}\|B_k\|^2} \lesssim C_d b_n^{\alpha-1} + C_d b_n^{-1/2} n^{\alpha/2}. \tag{27}$$

Thus, using (23), (24) and (27), we obtain

$$\mathbb{E} \left\| a_n^{-1} \sum_{k=1}^{a_n} b_n \bar{U}_k \bar{U}_k^\top - A^{-1} S A^{-1} \right\|$$

$$\lesssim \mathbb{E} \left\| a_n^{-1} \sum_{k=1}^{a_n} b_n^{-1} A_k A_k^\top - A^{-1} S A^{-1} \right\| + 2 a_n^{-1} \sum_{k=1}^{a_n} b_n^{-1} \left( \mathbb{E} \|A_k\| \|B_k\| + \mathbb{E} \|B_k\|^2 \right)$$

$$\lesssim C_d^2 n^{-\alpha/2 + (\beta-1)/4} + C_d^3 n^{-\alpha} + C_d a_n^{-1/2} + C_d b_n^{\alpha-1} + C_d b_n^{-1/2} n^{\alpha/2}.$$

□

**Lemma 4.** *Under the assumptions (A1), (A2) and (A3), and for sufficiently large $n$, we have*

$$\mathbb{E} \left\| a_n^{-1} \sum_{k=1}^{a_n} b_n (\bar{U}_k - \bar{U}_n)(\bar{U}_k - \bar{U}_n)^\top - A^{-1} S A^{-1} \right\|$$

$$\lesssim C_d^2 n^{-\alpha/2 + (\beta-1)/4} + C_d^3 n^{-\alpha} + C_d a_n^{-1/2} + C_d b_n^{\alpha-1} + C_d b_n^{-1/2} n^{\alpha/2} + C_d a_n^{-1}.$$

*Proof.* Observe that

$$a_n^{-1} \sum_{k=1}^{a_n} b_n (\bar{U}_k - \bar{U}_n)(\bar{U}_k - \bar{U}_n)^\top = a_n^{-1} \sum_{k=1}^{a_n} b_n \bar{U}_k \bar{U}_k^\top - a_n^{-1} \sum_{k=1}^{a_n} b_n \bar{U}_n \bar{U}_n^\top$$

$$= a_n^{-1} \sum_{k=1}^{a_n} b_n \bar{U}_k \bar{U}_k^\top - b_n \bar{U}_n \bar{U}_n^\top. \tag{28}$$

From Lemma 3, we have

$$\mathbb{E} \left\| a_n^{-1} \sum_{k=1}^{a_n} b_n \bar{U}_k \bar{U}_k^\top - A^{-1} S A^{-1} \right\| \lesssim C_d^2 n^{-\alpha/2 + (\beta-1)/4} + C_d^3 n^{-\alpha} + C_d a_n^{-1/2} + C_d b_n^{\alpha-1} + C_d b_n^{-1/2} n^{\alpha/2}. \tag{29}$$

Next, using

$$\bar{U}_n = n^{-1} S_0^{\tau_{a_n}} U_0 + n^{-1} \sum_{i=1}^{\tau_{a_n}} \left( S_i^{\tau_{a_n}} + I \right) \eta_i \xi_i,$$

and martingale property of $\xi_i$, we get

$$\mathbb{E}\|\bar{U}_n\|^2$$
$$\leqslant n^{-2}\|S_0^{\tau_{a_n}}\|^2\mathbb{E}\|U_0\|^2 + n^{-2}\sum_{i=1}^{\tau_{a_n}}\|\left(S_i^{\tau_{a_n}}+I\right)\|^2\eta_i^2\mathbb{E}\|\xi_i\|^2$$
$$\leqslant n^{-2}\|S_0^{\tau_{a_n}}\|^2\mathbb{E}\|U_0\|^2 + n^{-2}\sum_{i=1}^{\tau_{a_n}}\eta_i^2\|\left(S_i^{\tau_{a_n}}+I\right)\|^2[\text{tr}(S)+\mathbb{E}(\sigma_1\|D_i\|+\sigma_2\|D_i\|^2)]. \qquad (30)$$

Further, using Result 1 and Lemma 2, we have $\|S_0^{\tau_{a_n}}\|^2\mathbb{E}\|U_0\|^2 \lesssim C_d$, $\eta_i^2\|\left(S_i^{\tau_{a_n}}+I\right)\|$ is uniformly bounded, and $[\text{tr}(S)+\mathbb{E}(\sigma_1\|D_i\|+\sigma_2\|D_i\|^2)] \lesssim C_d$. Therefore,

$$n^{-2}\sum_{i=1}^{\tau_{a_n}}\eta_i^2\|\left(S_i^{\tau_{a_n}}+I\right)\|^2[\text{tr}(S)+\mathbb{E}(\sigma_1\|D_i\|+\sigma_2\|D_i\|^2)] \lesssim n^{-1}C_d. \qquad (31)$$

Hence, using (28)-(31), we get

$$\mathbb{E}\left\|a_n^{-1}\sum_{k=1}^{a_n}b_n(\bar{U}_k-\bar{U}_n)(\bar{U}_k-\bar{U}_n)^\top - A^{-1}SA^{-1}\right\|$$
$$\leqslant \mathbb{E}\left\|a_n^{-1}\sum_{k=1}^{a_n}b_n\bar{U}_k\bar{U}_k^\top - A^{-1}SA^{-1}\right\| + \mathbb{E}\left\|b_n\bar{U}_n\bar{U}_n^\top\right\|$$
$$\lesssim C_d^2 n^{-\alpha/2+(\beta-1)/4} + C_d^3 n^{-\alpha} + C_d a_n^{-1/2} + C_d b_n^{\alpha-1} + C_d b_n^{-1/2} n^{\alpha/2} + C_d a_n^{-1}.$$

□

## C  Consistency of batch-means estimator

We denote overall mean and batch means of $D_i$ as follows:

$$\bar{D}_n = \frac{1}{\tau_{a_n}}\sum_{i=1}^{\tau_{a_n}}D_i \text{ and } \bar{D}_k = \frac{1}{b_n}\sum_{i=\tau_{k-1}+1}^{\tau_k}D_i \text{ for } k=1,2,\ldots,a_n.$$

**Proof of Theorem 1.** We have the linear auxiliary iterate sequence $U_i = (I_d - \eta_i A)U_{i-1} +$

$\eta_i \xi_i$, $U_0 = D_0$. Let $\delta_i := D_i - U_i$, then

$$\delta_i = (I - \eta_i A)\delta_{i-1} + \eta_i(AD_{i-1} - \nabla F(\theta_{i-1})).$$

Define overall mean and batch means of $\delta_i$ as follows:

$$\bar{\delta}_n = \frac{1}{\tau_{a_n}} \sum_{i=1}^{\tau_{a_n}} \delta_i \text{ and } \bar{\delta}_k = \frac{1}{b_n} \sum_{i=\tau_{k-1}+1}^{\tau_k} \delta_i \text{ for } k = 1, 2, \ldots, a_n.$$

Now observe that

$$a_n^{-1} \sum_{k=1}^{a_n} b_n(\bar{\theta}_k - \bar{\theta}_n)(\bar{\theta}_k - \bar{\theta}_n)^\top$$

$$= a_n^{-1} \sum_{k=1}^{a_n} b_n \bar{D}_k \bar{D}_k^\top - a_n^{-1} \sum_{k=1}^{a_n} b_n \bar{D}_n \bar{D}_n^\top$$

$$= a_n^{-1} \sum_{k=1}^{a_n} b_n(\bar{U}_k + \bar{\delta}_k)(\bar{U}_k + \bar{\delta}_k)^\top - a_n^{-1} \sum_{k=1}^{a_n} b_n(\bar{U}_n + \bar{\delta}_n)(\bar{U}_n + \bar{\delta}_n)^\top$$

$$= a_n^{-1} \sum_{k=1}^{a_n} b_n(\bar{U}_k - \bar{U}_n)(\bar{U}_k - \bar{U}_n)^\top + a_n^{-1} \sum_{k=1}^{a_n} b_n(\bar{U}_k - \bar{U}_n)(\bar{\delta}_k - \bar{\delta}_n)^\top$$

$$+ a_n^{-1} \sum_{k=1}^{a_n} b_n(\bar{\delta}_k - \bar{\delta}_n)(\bar{U}_k - \bar{U}_n)^\top + a_n^{-1} \sum_{k=1}^{a_n} b_n(\bar{\delta}_k - \bar{\delta}_n)(\bar{\delta}_k - \bar{\delta}_n)^\top.$$

Then, by using $\|C\| \leq \operatorname{tr}(C)$ for a positive semidefinite matrix $C$, and Cauchy-Schwartz inequality, we have

$$\mathbb{E} \left\| a_n^{-1} \sum_{k=1}^{a_n} b_n(\bar{\theta}_k - \bar{\theta}_n)(\bar{\theta}_k - \bar{\theta}_n)^\top - A^{-1}SA^{-1} \right\|$$

$$\leq \mathbb{E} \left\| a_n^{-1} \sum_{k=1}^{a_n} b_n(\bar{U}_k - \bar{U}_n)(\bar{U}_k - \bar{U}_n)^\top - A^{-1}SA^{-1} \right\|$$

$$+ a_n^{-1} \sum_{k=1}^{a_n} b_n \mathbb{E} \operatorname{tr}[(\bar{\delta}_k - \bar{\delta}_n)(\bar{\delta}_k - \bar{\delta}_n)^\top]$$

$$+ \frac{2}{a_n} \sqrt{\sum_{k=1}^{a_n} b_n \mathbb{E} \operatorname{tr}[(\bar{U}_k - \bar{U}_n)(\bar{U}_k - \bar{U}_n)^\top] \sum_{k=1}^{a_n} b_n \mathbb{E} \operatorname{tr}[(\bar{\delta}_k - \bar{\delta}_n)(\bar{\delta}_k - \bar{\delta}_n)^\top]}, \quad (32)$$

and $\sum_{k=1}^{a_n} b_n \mathbb{E} \operatorname{tr}[(\bar{\delta}_k - \bar{\delta}_n)(\bar{\delta}_k - \bar{\delta}_n)^\top] \leq \sum_{k=1}^{a_n} b_n \mathbb{E} \operatorname{tr}[\bar{\delta}_k \bar{\delta}_k^\top]$.

35

Further, using the notations $Y_j^k$ and $S_j^k$ in Result 1, we get for $i > \tau_{k-1}$

$$\delta_i = (I - \eta_i A)\delta_{i-1} + \eta_i(AD_{i-1} - \nabla F(\theta_{i-1}))$$

$$= Y_{\tau_{k-1}}^i \delta_{\tau_{k-1}} + \sum_{j=\tau_{k-1}+1}^{i} Y_j^i \eta_j (AD_{j-1} - \nabla F(\theta_{j-1}))$$

$$\Rightarrow \bar{\delta}_k = \frac{1}{b_n} \sum_{i=\tau_{k-1}+1}^{\tau_k} Y_{\tau_{k-1}}^i \delta_{\tau_{k-1}} + \frac{1}{b_n} \sum_{i=\tau_{k-1}+1}^{\tau_k} \sum_{j=\tau_{k-1}+1}^{i} Y_j^i \eta_j (AD_{j-1} - \nabla F(\theta_{j-1}))$$

$$= \frac{1}{b_n} S_{\tau_{k-1}}^{\tau_k} \delta_{\tau_{k-1}} + \frac{1}{b_n} \sum_{j=\tau_{k-1}+1}^{\tau_k} \sum_{i=j}^{\tau_k} Y_j^i \eta_j (AD_{j-1} - \nabla F(\theta_{j-1}))$$

$$= \frac{1}{b_n} S_{\tau_{k-1}}^{\tau_k} \delta_{\tau_{k-1}} + \frac{1}{b_n} \sum_{j=\tau_{k-1}+1}^{\tau_k} (I + S_j^{\tau_k}) \eta_j (AD_{j-1} - \nabla F(\theta_{j-1})).$$

Now, by Cauchy Schwartz inequality,

$$\mathbb{E}\|\bar{\delta}_k\|^2 \leqslant \frac{2}{b_n^2} \|S_{\tau_{k-1}}^{\tau_k}\|^2 \mathbb{E}\|\delta_{\tau_{k-1}}\|^2 +$$

$$+ \frac{2}{b_n^2} \left( \sum_{j=\tau_{k-1}+1}^{\tau_k} \|(I + S_j^{\tau_k})\eta_j\|^2 \right) \mathbb{E} \left( \sum_{j=\tau_{k-1}+1}^{\tau_k} \|AD_{j-1} - \nabla F(\theta_{j-1})\|^2 \right). \quad (33)$$

Using Result 1 and Result 3, we have

$$\|S_{\tau_{k-1}}^{\tau_k}\|^2 \lesssim (\tau_{k-1} + 1)^{2\alpha}, \quad \sum_{j=\tau_{k-1}+1}^{\tau_k} \|S_j^{\tau_k} \eta_j\|^2 \lesssim b_n \quad (34)$$

$$\text{and} \quad \mathbb{E} \left( \sum_{j=\tau_{k-1}+1}^{\tau_k} \|AD_{j-1} - \nabla F(\theta_{j-1})\|^2 \right) \lesssim \sum_{j=\tau_{k-1}+1}^{\tau_k} L_F^2 \mathbb{E}\|D_{j-1}\|^4 \lesssim C_d^4 b_n \tau_{k-1}^{-2\alpha}. \quad (35)$$

Notice that Result 4 holds for both $U_i$ and $D_i$, so

$$\mathbb{E}\|\delta_{\tau_{k-1}}\|^2 \leqslant 2\mathbb{E}\|D_{\tau_{k-1}}\|^2 + 2\mathbb{E}\|U_{\tau_{k-1}}\|^2 \lesssim \tau_{k-1}^{-\alpha} C_d. \quad (36)$$

Therefore, using (33)-(36) and Lemma 1, we have

$$\mathbb{E}\|\bar{\delta}_k\|^2 \lesssim \frac{1}{b_n^2} \tau_{k-1}^{\alpha} + \frac{1}{b_n^2} b_n C_d^4 b_n \tau_{k-1}^{-2\alpha} \lesssim \frac{1}{b_n^2} \tau_{k-1}^{\alpha} + C_d^4 \tau_{k-1}^{-2\alpha}$$

$$\Rightarrow a_n^{-1} \sum_{k=1}^{a_n} b_n \mathbb{E}\|\bar{\delta}_k\|^2 \lesssim C_d^4 b_n a_n^{-1} \sum_{k=1}^{a_n} \tau_{k-1}^{-2\alpha} \lesssim C_d^4 b_n^{1-2\alpha} a_n^{-1} \sum_{k=1}^{a_n} k^{-2\alpha} \asymp C_d^4 b_n^{1-2\alpha} a_n^{-2\alpha} = C_d^4 n^{-2\alpha} b_n.$$

(37)

Thus, using (32) and (37), we obtain

$$\mathbb{E}\left\| a_n^{-1} \sum_{k=1}^{a_n} b_n (\bar{\theta}_k - \bar{\theta}_n)(\bar{\theta}_k - \bar{\theta}_n)^\top - A^{-1} S A^{-1} \right\|$$
$$\lesssim C_d^2 n^{-\alpha/2 + (\beta-1)/4} + C_d^3 n^{-\alpha} + C_d a_n^{-1/2} + C_d b_n^{\alpha-1} + C_d b_n^{-1/2} n^{\alpha/2} + C_d a_n^{-1} + C_d^4 n^{-2\alpha} b_n.$$

□

## D  Lugsail estimator

### D.1  Details for Example 1

The mean estimation model is given by

$$y = \theta^* + \epsilon,$$

where $\theta^* \in \mathbb{R}$ and $\epsilon$ is the random error term with mean zero. Let $y_i$ be a sequence of iid observations from the model. Consider the square error loss function $F(\theta) = (y-\theta)^2/2$. Without loss of generality take $\theta^* = 0$ and $\theta_0 = 0$, then the $i^{th}$ SGD iterate has the form

$$\theta_i = \theta_{i-1} + \eta_i(y_i - \theta_{i-1}) = (1 - \eta_i)\theta_{i-1} + \eta_i \epsilon_i, \ i \geqslant 1. \tag{38}$$

This implies

$$\theta_i = \sum_{p=1}^{i} \prod_{k=p+1}^{i} (1 - k^\alpha) p^{-\alpha} \epsilon_p. \tag{39}$$

Now, the estimand is

$$\text{Var}(\sqrt{n}\,\hat{\theta}_n) = n \text{ Var}\left(\frac{1}{a_n}\sum_{k=1}^{a_n}\bar{\theta}_k\right) = \frac{b_n^2}{n}\mathbb{E}\left(\sum_{k=1}^{a_n}\bar{\theta}_k\right)^2, \qquad (40)$$

and the proposed estimator is

$$\hat{\Sigma}_n = \frac{1}{a_n}\sum_{k=1}^{a_n} b_n(\bar{\theta}_k - \hat{\theta}_n)^2 = \frac{b_n^2}{n}\sum_{k=1}^{a_n}\bar{\theta}_k^2 + \frac{b_n}{n}O_p(1). \qquad (41)$$

Thus, ignoring the second term, which is tending to zero at a higher rate, we have

$$\begin{aligned}\text{Bias}(\hat{\Sigma}_n) &= \mathbb{E}(\hat{\Sigma}_n) - \text{Var}(\sqrt{n}\,\hat{\theta}_n) \\ &\simeq \frac{-2b_n^2}{n}\sum_{1\leqslant j<k\leqslant a_n} Cov(\bar{\theta}_j, \bar{\theta}_k) = \frac{-2}{n}\sum_{1\leqslant j<k\leqslant a_n} Cov(b_n\bar{\theta}_j, b_n\bar{\theta}_k). \end{aligned} \qquad (42)$$

Further, using the fact that $Cov(\theta_p, \theta_q) = C_1 q^{-\alpha}(1 - q^{-\alpha})^{q-p}$ for $p < q$, where $C_1$ is fixed constant, we get for $j < k$

$$\text{Cov}(b_n\bar{\theta}_j, b_n\bar{\theta}_k) = \sum_{p=\tau_{j-1}+1}^{\tau_j}\sum_{q=\tau_{k-1}+1}^{\tau_k} Cov(\theta_p, \theta_q) = C_1 \sum_{p=\tau_{j-1}+1}^{\tau_j}\sum_{q=\tau_{k-1}+1}^{\tau_k} q^{-\alpha}(1-q^{-\alpha})^{q-p}. \qquad (43)$$

Using (42) and (43), we get

$$\text{Bias}(\hat{\Sigma}_{b_n}) \approx \frac{-2}{n}\sum_{1\leqslant j<k\leqslant a_n} Cov(b_n\bar{\theta}_j, b_n\bar{\theta}_k) = \frac{-2C_1}{n}\sum_{1\leqslant j<k\leqslant a_n}\sum_{p=\tau_{j-1}+1}^{\tau_j}\sum_{q=\tau_{k-1}+1}^{\tau_k} q^{-\alpha}(1-q^{-\alpha})^{q-p}.$$

### D.2 Proof of Theorem 2

*Proof.* Observe that

$$\begin{aligned}&\mathbb{E}\|\hat{\Sigma}_{L,b_n} - A^{-1}SA^{-1}\| \\ &\leqslant \mathbb{E}\|2(\hat{\Sigma}_{n,2b_n} - A^{-1}SA^{-1}) - (\hat{\Sigma}_{b_n} - A^{-1}SA^{-1})\| \\ &\leqslant 2\mathbb{E}\|(\hat{\Sigma}_{n,2b_n} - A^{-1}SA^{-1})\| + \mathbb{E}\|(\hat{\Sigma}_{b_n} - A^{-1}SA^{-1})\| \\ &\lesssim 4\left(C_d^2 n^{-\alpha/2+(\beta-1)/4} + C_d^3 n^{-\alpha} + C_d\left(\frac{a_n}{2}\right)^{-1/2} + C_d(2b_n)^{\alpha-1} + C_d(2b_n)^{-1/2}n^{\alpha/2} + C_d\left(\frac{a_n}{2}\right)^{-1} + C_d^4 n^{-2\alpha}2b_n\right)\end{aligned}$$

38

$$+ \ C_d^2 n^{-\alpha/2+(\beta-1)/4} + C_d^3 n^{-\alpha} + C_d a_n^{-1/2} + C_d b_n^{\alpha-1} + C_d b_n^{-1/2} n^{\alpha/2} + C_d a_n^{-1} + C_d^4 n^{-2\alpha} b_n$$

$$\leqslant 5C_d^2 n^{-\alpha/2+(\beta-1)/4} + 5C_d^3 n^{-\alpha} + 6C_d a_n^{-1/2} + 5C_d b_n^{\alpha-1} + 4C_d b_n^{-1/2} n^{\alpha/2} + 9C_d a_n^{-1} + 9C_d^4 n^{-2\alpha} b_n.$$

$\square$


\end{document}